\documentclass[floatfix,twocolumn,superscriptaddress,amsmath,showpacs,aps,pre]{revtex4-1}
\usepackage{epsfig,color}

\newcommand{\eff}{\rm{eff}}

\begin{document}

\title{Langevin dynamics for vector variables driven by  multiplicative
white noise: a functional formalism}

\author{Miguel Vera Moreno}
\affiliation{Departamento de F{\'\i}sica Te\'orica,
Universidade do Estado do Rio de Janeiro, Rua S\~ao Francisco Xavier 524,
20550-013,  Rio de Janeiro, RJ, Brazil.}

\author{Zochil Gonz\'alez Arenas}
\affiliation{Departamento de F{\'\i}sica Te\'orica,
Universidade do Estado do Rio de Janeiro, Rua S\~ao Francisco Xavier 524,
20550-013,  Rio de Janeiro, RJ, Brazil.}

\author{Daniel G.\ Barci}
\affiliation{Departamento de F{\'\i}sica Te\'orica,
Universidade do Estado do Rio de Janeiro, Rua S\~ao Francisco Xavier 524,
20550-013,  Rio de Janeiro, RJ, Brazil.}

\date{December 22, 2014}

\begin{abstract}
We discuss  general multi-dimensional stochastic processes driven by a system of Langevin equations with  multiplicative white noise. In particular,  we address the problem of how time reversal diffusion processes are affected by the variety of conventions available to deal with stochastic integrals.  
We  present a functional formalism to built up the generating functional of correlation functions without any  type of discretization  of the Langevin equations at any intermediate step. The generating functional is characterized by a functional integration over two sets of commuting variables as well as Grassmann variables. In this representation, time reversal transformation became a linear transformation in the extended variables, simplifying in this way the complexity introduced by the mixture of prescriptions and the associated calculus rules.  The stochastic calculus is codified in our formalism in the structure of the Grassmann algebra. 
We study some examples such as higher order derivatives Langevin equations and the functional representation of  the micromagnetic stochastic Landau-Lifshitz-Gilbert equation. 
\end{abstract}

%%%%%%%%%%%%%%%  PACS %%%%%%%%%%%%%%%%%%%%%%%%%%%%%%
%05.40.-a 	Fluctuation phenomena, random processes, noise, and Brownian motion (for fluctuations in superconductivity, see 74.40.-n; for statistical theory and fluctuations in nuclear reactions, see 24.60.-k; for fluctuations in plasma, see 52.25.Gj; for nonlinear dynamics and chaos, see 05.45.-a)

%02.50.Ey 	Stochastic processes

%05.10.Gg 	Stochastic analysis methods (Fokker-Planck, Langevin, etc.) 

%02.50.Ga 	Markov processes 
%%%%%%%%%%%%%%%%%%%%%%%%%%%%%%%%%%%%%%%%%%%%%%%%%%

\pacs{05.40.-a, 02.50.Ey, 05.10.Gg, 02.50.Ga}

\maketitle

%%%%%%%%%%%%%%%%
\section{Introduction}
%%%%%%%%%%%%%%%%

Out of equilibrium statistical mechanics is still a major topical subject due to the difficulty in the formulation of a general  closed theory at the same level of equilibrium statistical mechanics. For this reason, there are many different approaches to study this type of systems.  
Stochastic dynamics\cite{gardiner,vanKampen} provides an interesting approach to
out of equilibrium statistical mechanics~\cite{crooks1999,seifert2005}.
Interestingly, it is possible to attribute thermodynamical concepts like heat,
entropy or free energy to each trajectory of an stochastic evolution, giving rise
to the research field usually called  stochastic
thermodynamics~\cite{seifert2008}. 

There are many different ways to deal with stochastic processes and its applications can be found along a wide area of scientific research. Stochastic differential equations (SDE), such as a system of Langevin equations, could be one of the most popularly used approaches.  An equivalent formalism is the  Fokker-Planck equation, a partial differential equation for the time dependent probability density. The path integral formulation provides an alternative formalism, where a probability density is assigned to each stochastic trajectory and it is very similar to the development of Feynman path integrals in quantum mechanics \cite{MSR1973,Janssen1976,deDominicis}.  

The theory of stochastic evolution brings a beautiful connection 
between dynamics and statistical physics. In general, Einstein
relation or, more generally, fluctuation-dissipation relations associate
dynamical properties with thermodynamical equilibrium.
However, stochastic dynamics does not necessarily model physical systems~\cite{Poschel,Murray,Mantegna,Bouchaud}, where
the long time evolution should conduce to thermodynamical equilibrium. In fact,
it is possible to have more general stationary-state distributions which
represent  equilibrium states in a stochastic dynamical sense, not
related to thermodynamics.

Typically, a Langevin equation describes the motion of a diffusive particle in a medium 
which is modeled by splitting its effects in two parts: a deterministic part, given by an homogeneous viscous
force, and a stochastic one, given by a random force with zero expectation value. In this way, fluctuations enter as an {\em additive} noise.  
However, the viscous force could have non-homogeneous
contributions, for instance, in the presence of boundary conditions, such as
a diffusion  of a Brownian particle near a wall~\cite{Lancon2001,Lancon2002}.
In  the case of  non-homogeneous diffusion, fluctuations depend on the
state of the system,  defining a {\em multiplicative} stochastic process.  An 
interesting example of multiplicative noise is the stochastic
Landau-Lifshitz-Gilbert equation~\cite{Palacios1998,Bertottl},
used to describe dynamics of classical magnetic moments of individual
nanoparticles. In this case, the noisy fluctuations of the magnetic field
couple the magnetic moment in a multiplicative way.  

In this paper, we are interested in the path integral formulation of multiplicative stochastic processes. This is an interesting topic as long as the path integral formalism provides a useful technique to compute correlation and response functions. Once a generating functional for correlation functions is built, it would be possible to study the system using all the machinery developed in quantum field theory, even non-perturbative techniques. The relation between symmetries and fluctuation theorems is also a very interesting issue that can be face with this technique. 
The path integral formulation of multiplicative noise processes has been studied before~\cite{Janssen-RG}. In particular, a generating functional for colored noise processes was presented in Ref.~\cite{AronLeticia2010}. However, the special case of Markov processes with Gaussian white noise is more subtle due to the variety of stochastic prescriptions available to define the Wiener integrals. In this sense, the main difficulty is given by the fact that each prescription represents a different process and implies specific rules of calculus, which does not obey, in particular, the traditional chain rule (except in the Stratonovich case). We have previously presented a set of results in this  subject~\cite{arenas2010,arenas2012,Arenas2012-2} for the case of a  single stochastic variable. Using a path integral formalism we have shown that, even in the case of multiplicative noise, there is a hidden supersymmetry that encodes the equilibrium properties of the system. 

In the present paper, we generalize the formalism of Ref.~\cite{arenas2012,Arenas2012-2} to the case of multiple variable systems.  
While some formal aspects of the generalization are straightforward, we point out important issues specific of multiple variable systems, which are not present in  single variable ones. 

We consider a vector stochastic variable whose dynamics is driven  by  a system of Langevin equations with multiplicative white noises. 
We discuss the concept of time reversibility and carefully deduce the time reversal transformation for a multivariate diffusion process. 
We perform a  derivation of the generating functional, without discretizing the Langevin equations at any intermediate step. To do that, we  use auxiliary  Grassmann variables, which allow us to describe  a stochastic multiplicative process without any reference to the definition of  Wiener integrals. The known prescriptions which define the stochastic process (It\^o, Stratonovich, etc.), emerged in our formalism in the definition of Green's functions at definite points.  

In order to illustrate our general method, we discuss two specific applications. We discuss the case of non-Markov processes driven by higher derivative order Langevin equations, and we also show an application to the  Stochastic Landau-Lifshitz-Gilbert equation for micro-magnetic dynamics.  Through theses examples, it is interesting to note, how the intricacies  of the stochastic calculus are compactly codified in the Grassmann algebra in many  different situations.   

We have also included an appendix with a detailed demonstration
of the main stochastic calculus formulae using the Generalized Stratonovich prescription.  In particular we carefully discuss the \emph{generalized chain rule}, 
 used without demonstration in Refs.~\cite{arenas2012,Arenas2012-2,Aron2014} and  we extend it for the multivariate case.

The paper has the following structure. In the next Section we describe our model, consisting in a system of stochastic differential equations with a general vector drift and a general state-dependent diffusion matrix. We pay special attention to the stochastic prescription which complete the definition of the differential equations and in the stochastic calculus induced by these definitions. 
In Sec.~\ref{sec:TimeReversal} we discuss  time reversal diffusion, describing in detail how a time reversal transformation depends on the stochastic prescription and how it is implemented in the trajectory space.  In Sec.~\ref{sec:GeneratingFunctional} we built up the generating functional of correlation functions in terms of a set of commuting variables as well as Grassmann variables.  Then, in Sec.~\ref{sec:GrassmanIntegration}, we explicitly integrate over Grassmann variables to show how the specific stochastic prescription appears as a  definition of Green's functions at the origin. We develop various examples in Sec.~\ref{sec:Examples} and we finally discuss our results in Sec.~\ref{sec:discusion}.  We dedicate two appendices to some important mathematical details. In  appendix~\ref{sec:alphacalculus} we present a summary of stochastic calculus in the generalized Stratonovich prescription. In particular, we present an explicit demonstration of the generalized chain-rule. Moreover, in appendix~\ref{sec:Determinants} we summarize properties of Grassmann algebra and determinant representations. 

%%%%%%%%%%%%%%%%%%%%%%%%%%%%%%
\section{Langevin equations with multiplicative noise}
\label{sec:Langevin}
%%%%%%%%%%%%%%%%%%%%%%%%%%%%%%
We consider the system of Langevin equations:
\begin{equation}
 \frac{dx_i(t)}{dt} = f_i({\bf x}(t)) + 
g_{ij}({\bf x}(t))\eta_j(t)   \label{eq:LangevSystem}
\end{equation}
 where  $i = 1,\ldots,n$, $j = 1,\ldots,m$, ${\bf x} \in \Re^n$ and  $\eta_j(t)$ are $m$ independent Gaussian white noises,
\begin{equation}
 \left\langle \eta_i(t)\right\rangle   = 0 \;\;\mbox{,}\;\; 
\left\langle  \eta_i(t), \eta_j(t')\right\rangle = \delta_{ij} \delta(t-t').
\label{whitenoise}
\end{equation}
We use bold face characters for vector variables and  summation over repeated indices is understood. 
The drift force $f_i({\bf x})$ and  the diffusion matrix $g_{ij}({\bf x})$ are,  in principle,  arbitrary smooth functions of ${\bf x}(t)$. This kind of equations is 
commonly used to describe many diffusion processes occurring in nature under some random influence, where $x_i$ represents a physical macroscopic quantity which undergoes diffusion. In the single-variable case, Eq.~\eqref{eq:LangevSystem} describes a Markov stochastic process. However, in the multidimensional case, the process can be Markovian or not, depending on the structure of the diffusion matrix $g_{ij}$. Some examples of this situation are developed in Section~\ref{sec:Examples}.

\begin{figure}[tbp]
\includegraphics[height=5.8 cm]{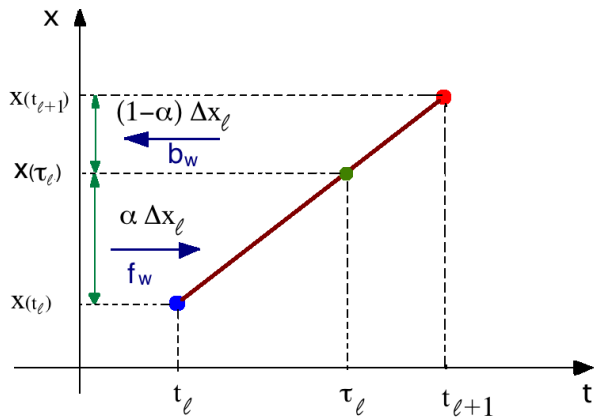}
\caption{Generalized Stratonovich prescription}
\label{fig:AlphaPrescription}
\end{figure}

As is well known, due to the delta-correlated nature of $\eta_j(t)$, it is necessary to give sense to the ill-defined products $g_{ij}({\bf x}(t))\eta_j(t)$ to completely define Eq.~(\ref{eq:LangevSystem}). 
The problem can be visualized by looking at the integral
\begin{equation}
\int    g_{ij}({\bf x}(t))\; \eta_j(t) dt= \int  g_{ij}({\bf x}(t))\;  dW_j(t) \ ,
\end{equation} 
where we have defined Wiener processes $W_j(t)=\int_{t_0}^t \eta_j(t') dt'$. 
By definition, the Riemann-Stieltjes integral is given by 
\begin{eqnarray}
\lefteqn{
\int   g_{ij}({\bf x}(t))\;  dW_j(t)= }\nonumber  \\
&=& \lim _{n\to\infty} \sum_{\ell=1}^n  g_{ij}({\bf x}(\tau_\ell))(W_j(t_{\ell})-W_j(t_\ell-1)),
\label{eq.Wiener}
\end{eqnarray}
where the time integration domain have been discretized and $\tau_\ell$ is taken in the interval $[t_{\ell-1},t_\ell]$ with $\ell=1,\ldots,n$. The limit above is taken
in the sense of the {\em mean-square limit} (see Appendix \ref{sec:ItoFormula}). For a smooth measure
$W(t)$, the limit converges to a unique value, regardless the value of
$\tau_\ell$. 
However, $W(t)$ is not smooth,  in fact, it is nowhere differentiable. In any interval, the white noise fluctuates an infinite number of times with infinite variance. 
Moreover, the value of the integral depends on the place $\tau_\ell$ is chosen in the interval $[t_{\ell-1},t_\ell]$, which is the origin of the variety of stochastic prescriptions available to define this integral. At the same time, some aspects related with this decision are the core of what is known as the \emph{It\^o -- Stratonovich dilemma}. 
All these prescriptions can be summarized in the so-called ``generalized Stratonovich prescription''~\cite{Hanggi1978} 
or ``$\alpha$-prescription''~\cite{Janssen-RG}, for which we choose 
\begin{equation}
g_{ij}(x(\tau_\ell))=g_{ij}(\alpha x(t_\ell) + (1-\alpha)x(t_{\ell-1})) \;,  
\label{eq.prescription}
\end{equation}
 with $0\le \alpha \le 1$.
In figure~(\ref{fig:AlphaPrescription}) we sketch this prescription showing the interval $[t_{\ell-1},t_\ell]$ and clearly indicating the linear interpolation between the pre-point value $\alpha=0$ and the post-point one, $\alpha=1$.  
In this way, $\alpha=0$ corresponds with the pre-point It\^o interpretation and $\alpha=1/2$ coincides with the (mid-point) Stratonovich one.  
Moreover, the post-point prescription, $\alpha=1$, is also known as the kinetic or the H\"anggi-Klimontovich
interpretation~\cite{Hanggi1978, Hanggi1980, Hanggi1982, Klimontovich}.

In principle, each particular choice of $\alpha$ fixes a different stochastic evolution and, once the interpretation is fixed, the stochastic dynamics is unambiguously defined. So, in order to  completely define the stochastic 
process described by Eq.~(\ref{eq:LangevSystem}) we need to fix, not only the  set of functions given by the drift  $f_i$ and the diffusion matrix $g_{ij}$, but also  the parameter $0\le\alpha\le 1$.

Fortunately, any stochastic process described by a SDE in the $\alpha$-prescription, can also be described by {\it another SDE} in any other prescription $\beta$.
To be precise, suppose that ${\bf x(t)}$ is a solution of the system~\eqref{eq:LangevSystem} of SDE, where the Wiener integral is understood in the  $\alpha$-prescription. Then, ${\bf x(t)}$ is also a solution of  
\begin{equation}
 \frac{dx_i}{dt} = \left[f_i({\bf x})+(\alpha-\beta)g_{kj}({\bf x})\partial_k g_{ij}({\bf x}) \right] + 
g_{ij}({\bf x})\eta_j \ , 
\label{eq:LangevSystemBeta}
\end{equation}
interpreted in the $\beta$-prescription.  In Eq.~(\ref{eq:LangevSystemBeta}) and in the rest of the paper 
$\partial_k\equiv \partial/\partial x_k$.
Therefore, we can represent the {\em same stochastic process} in different prescriptions by just shifting the drift through
\begin{equation}
f_i({\bf x})\to f_i({\bf x})+(\alpha-\beta)  g_{k\ell}({\bf x})\partial_k g_{i\ell}({\bf x}).
\label{eq:Translator}
\end{equation}
A detailed demonstration of Eq.~(\ref{eq:Translator}) can be found in appendix~(\ref{Ap:alphabeta}).
Note that there is a special condition on the diffusion matrix in which the stochastic evolution is independent of the particular prescription used. If, for instance~\cite{Hanggi1978}, 
\begin{equation}
g_{k\ell}({\bf x})\partial_k g_{i\ell}({\bf x})=0,
\label{eq:gdg}
\end{equation}
then the evolution is independent of the prescription. We will show  in section~\ref{sec:Examples} two particular examples of this relation. 

An important observation is that different values of $\alpha$ imply different rules of calculus. In fact, the Stratonovich prescription, $\alpha=1/2$, is the only one where the rules of calculus are the usual ones.  We can summarize this statement showing the 
``chain rule'' in the $\alpha$-prescription. Consider, for instance, an arbitrary function of the stochastic variable ${\bf x}(t)$ satisfying the Eq.~(\ref{eq:LangevSystem}) in the $\alpha$-prescription. We can show (see appendix~\ref{Ap:chainrule1} and~\ref{Ap:chainrule2} for a rigorous demonstration) that 
\begin{equation}
\frac{dF({\bf x}(t))}{dt}=\partial_k F \frac{dx_k}{dt}+\left(\frac{1-2\alpha}{2}\right)g_{ik}g_{jk}\partial_i\partial_jF.
\label{CR}
\end{equation}
Notice that,  by fixing $\alpha=1/2$, Eq.~(\ref{CR}) is the chain rule of the usual calculus. However, 
for $\alpha\neq 1/2$, the chain rule gets an extra term proportional to the square of the dissipation matrix. 
Of course, this affects all type of calculations rules like, for instance, integration by parts. 

Therefore, once the stochastic process is uniquely determined, we can formulate the description (using the drift transformation, Eq.~(\ref{eq:Translator})) in terms of a SDE using any convention.  
The final convention used in the calculations is a matter of taste. 
For instance, It\^o prescription is more suitable for numerical computation, while the Stratonovich one is preferably used for analytic calculations, due to its timely property of preserving the usual rules of calculus. However, it is convenient to formulate a prescription-independent formalism as long as some properties, like time reversibility, mixes different prescriptions. In the next section we discuss this point in some detail.

%%%%%%%%%%%%%%%%%%%%%%%%%%%
\section{Time reversal diffusion}
\label{sec:TimeReversal}
%%%%%%%%%%%%%%%%%%%%%%%%%%%
Time reversal properties of diffusion processes play a central role in statistical mechanics in equilibrium as well as in  out-of-equilibrium systems. However, the concept of time reversal in a diffusion process is tricky. In a multiplicative process driven by white noise, time reversal is a very subtle issue since the definition of backward trajectories depends on the stochastic prescription $\alpha$.  
In fact, if we consider a single stochastic variable $x(t)$
whose time evolution represents a Markov diffusion process described by an It\^o SDE,  then it is possible to demonstrate~\cite{Haussmann1986,Millet1989} that the  time reversed evolution is also a Markov diffusion process, however with a different drift.   

\begin{figure}
\centering
\includegraphics[height=6.5cm]{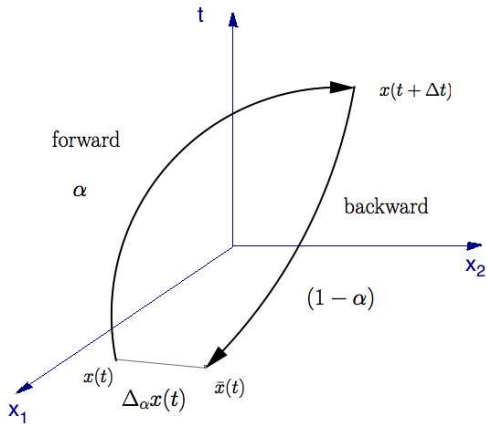}
\caption{Sketch of a ``time-loop''  evolution of Langevin Eq.~(\ref{eq:LangevSystem}). The forward evolution $t\to t+\Delta t$ is performed with a
prescription $\alpha$,  while the backward trajectory $t+\Delta t \to t$ evolves
with the time reversal conjugate prescription $(1-\alpha)$. Except for
the Stratonovich convention  $\alpha=1/2$, $\Delta x(t)_\alpha\neq 0$.}
 \label{fig:timereversal}
 \end{figure}

In this section, we study  this point in more detail for a vector variable ${\bf x}(t)$ satisfying a  system of Langevin SDE in the $\alpha$-prescription. We focus on how to correctly define the stochastic trajectories in order to apply this concept in a path integral formalism.  
Consider, for simplicity,  the system of Langevin Eqs.~(\ref{eq:LangevSystem}), in the simplest case of   $f_i({\bf x})=0$. Naively, we could compute the backward time evolution of 
${\bf x}(t)$ by just changing the sign on the velocity  $d{\bf x}/dt\to -d{\bf x}/dt$, in such
a way that, for a given noise configuration,  the particle turns back on its
own feet.  Looking closely, we immediately realize that the problem is more complex.
Consider, for instance, a time interval $(t, t+\Delta t)$. The forward evolution is obtained
by integrating Eq.~(\ref{eq:LangevSystem}) (with $f_i=0$) between the initial and final times
$t$ and  $t+\Delta t$, respectively,   
\begin{equation}
x_i(t+\Delta t )=x_i(t)+ \int_t^{t+\Delta t}g_{ij}({\bf x})dW_j;
\label{eq.fwevolution}
\end{equation}
while the backward evolution, considering $x_i(t+\Delta t )$ as the initial condition,   is simply obtained by
\begin{equation}
\bar x_i(t)=x_i(t+\Delta t )+ \int^t_{t+\Delta t}g_{ij}({\bf x})dW_j,
\label{eq.bwevolution}
\end{equation}
where we use the notation $\bar {\bf x}$ just to differentiate backward from forward
evolution. If the integrals were ``normal'' integrals, we could use the trivial property $\int_a^b =-\int_b^a$. In that case, there would be no 
need to differentiate backward and forward variables, since
Eqs.~(\ref{eq.fwevolution}) and~(\ref{eq.bwevolution}) would be the same 
equation and ${\bf x}(t)=\bar {\bf x}(t)$. 
However, the integrals are Wiener integrals and need to be carefully defined.
As we saw in the last section, 
\begin{eqnarray}
\lefteqn{
\int_t^{t+\Delta t}    g_{ij}({\bf x}(t))\;  dW_j(t)= } \nonumber \\
&&\lim _{n\to\infty} \sum_{k=1}^n  g_{ij}({\bf x}(\tau_k))(W_j(t_{k})-W_j(t_{k-1})),
\label{eq.Wienerfw}
\end{eqnarray}
with  
\begin{equation}
g({\bf x}(\tau_k))=g(\alpha{\bf x}(t_k)+ (1-\alpha) {\bf x}(t_{k-1})), \ \ \ 0\le \alpha \le 1. 
\label{eq.prescriptionfw}
\end{equation}
The time reversal integral is obtained by changing $t_{k-1}\leftrightarrow t_k$
in Eq.~(\ref{eq.Wienerfw}). The important point is  that 
$g(x(\tau_k))$ also depends on $(t_{k-1},t_{k})$.
Then,
\begin{eqnarray}
\lefteqn{
\int^t_{t+\Delta t}  g_{ij}({\bf x}(t))\;  dW_j(t)=} \nonumber \\
&& \lim _{n\to\infty} \sum_{k=1}^n  \bar g_{ij}({\bf x}(\tau_k))(W_j(t_{k-1}) - W_j(t_{k})),
\label{eq.Wienerbw}
\end{eqnarray}
with  
\begin{equation}
\bar g({\bf x}(\tau_k))=g(\alpha {\bf x}(t_{k-1}) + (1-\alpha) {\bf x}(t_{k})), \ \ \ 0\le \alpha \le 1,
\label{eq.prescriptionbw}
\end{equation}
where $\bar g({\bf x})$ was obtained from Eq.~(\ref{eq.prescriptionfw}), by
replacing $t_{k-1}\leftrightarrow t_{k}$ or, equivalently, $\alpha \to (1-\alpha)$.
This fact can be intuitively understood from figure~(\ref{fig:AlphaPrescription}). The arrows inside the figure indicate the forward ($f_w$) and backward ($b_w$) time directions. We see, for instance, that the pre-point prescription in the forward direction $\alpha=0$ is, in fact, the post-point prescription in the backward direction. 
Therefore, the time reversed stochastic evolution is characterized by the
transformations 
\begin{equation}
{\bf x}(t)\to {\bf x}(-t)  ~~~~~\mbox{and}~~~~~~~\alpha \to (1-\alpha).
\end{equation}
In this sense, we say that the prescription $(1-\alpha)$ is the time reversal
conjugate of $\alpha$. Then, the post-point H\"anggi-Klimontovich
interpretation (also known as anti-It\^o prescription) is the time reversal conjugate of the pre-point It\^o one, and vice versa. The only
{\em time reversal invariant prescription} is the Stratonovich one,
$\alpha=1/2$. 
This means that, except for the Stratonovich case, the backward and forward
stochastic paths do not have the same end points. This is illustrated in figure~(\ref{fig:timereversal}), 
where we compute a ``time-loop'' evolution. Consider we
want to compute the evolution of the system starting at ${\bf x}(t)$, going forward a
time interval $\Delta t$ and then, turning back in time the same interval
$-\Delta t$.
Using  Eqs.~(\ref{eq.fwevolution}) and~(\ref{eq.bwevolution})  and taking into account Eqs.~(\ref{eq.Wienerfw}) and~(\ref{eq.Wienerbw}), it is immediate to compute, 
\begin{eqnarray}
\Delta_\alpha x_i(t)&\equiv& \bar x_i(t)-x_i(t) \nonumber  \\
&=& \int_t^{t+\Delta t}g^{(\alpha)}_{ij} dW_j- \int_t^{t+\Delta t}
g^{(1-\alpha)}_{ij} dW_j , 
\label{eq:deltax}
\end{eqnarray}
where we have explicitly indicated that the first integral should be taken in the $\alpha$-prescription while the second one is in the $(1-\alpha)$-interpretation.
To explicitly compute $\Delta_\alpha {\bf x}(t)$, we use Eq.~(\ref{eq:Translator}) to write both integrals in the same prescription. Setting $\beta = 1-\alpha$ in Eq.~(\ref{eq:Translator}) we find 
\begin{eqnarray}
\int_t^{t+\Delta t}g^{\alpha}_{ij}dW_j&=&\int_t^{t+\Delta t}g^{1-\alpha}_{ij}dW_j 
\nonumber \\
&+& (2\alpha-1) \int_t^{t+\Delta t}  dt'\; g_{k\ell}  \partial_k  g_{i\ell}.    
\end{eqnarray}
Replacing  this expression into Eq.~(\ref{eq:deltax}) and con\-si\-de\-ring a very small 
$\Delta t$, we immediately find 
\begin{equation}
\Delta_\alpha x_i(t) =(2\alpha-1)g_{k\ell}({\bf x})  \partial_k  g_{i\ell}({\bf x}) \Delta t.
\label{eq:DeltaXalpha}
\end{equation}
We see that, in general, the forward and backward time evolutions do not have the same endpoints. 
The only exception is the Stratonovich prescription $\alpha=1/2$ and the special case of 
 $g_{k\ell}({\bf x})  \partial_k  g_{i\ell}({\bf x})=0$, where $\Delta_\alpha {\bf x}(t)=0$, in both cases, and the loops are closed. 
The fact that  $\Delta_\alpha {\bf x}(t)\neq 0$, also implies that the forward and backward time evolution cannot be described by the same drift function.

%%%%%%%%%%%%%%%%%%%%%%%%
%%%%%%%%%%%%%%%%%%%%%%%%

In order to address this point we find more convenient to work in the Fokker-Planck formalism.  The time dependent probability $P({\bf x}, t)$ for 
the vector variable ${\bf x}$ satisfying Eq.~(\ref{eq:LangevSystem}), fulfills the following partial differential equation~\cite{Lubensky2007} 
\begin{equation}
\frac{\partial P({\bf x}, t)}{\partial t}+ \nabla \cdot  {\bf J}({\bf x},t)=0,
\label{eq:FP}
\end{equation}
where the current ${\bf J}({\bf x},t)$ is given by 
\begin{equation}
J_i({\bf x},t)=[f_i+\alpha g_{k\ell}\partial_k g_{i\ell}]P({\bf x}, t) -\frac{1}{2}\partial_j[g_{i\ell}g_{j\ell}P({\bf x}, t)].
\label{eq:Current}
\end{equation}
Clearly, the solution of Eq.~(\ref{eq:FP}), with~(\ref{eq:Current}), depends on the parameter $\alpha$, as anticipated. 
For reasonable functions $f_i({\bf x})$ and $g_{ij}(\bf x)$, we expect that, at very long times, the probability converges to a stationary state 
\begin{equation}
P_{\rm st}({\bf x})=\lim_{t\to+\infty} P({\bf x},t ),
\end{equation}
 for which, 
\begin{equation}
\nabla \cdot{\bf J}_{\rm st}({\bf x})= 0.
\end{equation}
For an equilibrium state we have ${\bf J}_{\rm st}=0$, but for a general out-of-equilibrium stationary state, it is sufficient to have a divergence-less current. 

We can write the Fokker-Plank equation for a backward trajectory by replacing $t\to -t$, $\alpha\to 1-\alpha$ and 
${\bf f}\to \hat{\bf f}$. Then, the backward probability $\hat P({\bf x}, t)$ satisfies 
\begin{equation}
\frac{\partial \hat P({\bf x}, t)}{\partial t}+ \nabla \cdot  {\hat{{\bf J}}}({\bf x},t)=0,
\end{equation}
where
\begin{equation}
\hat J_i({\bf x},t)=-[\hat f_i+(1-\alpha) g_{k\ell}\partial_k g_{i\ell}]\hat P+\frac{1}{2}\partial_j(g_{i\ell}g_{j\ell}\hat P).
\label{eq:CurrentBackward}
\end{equation}
Evidently, forward and backward evolutions of the probability are quite different. However, the stationary states of both equations should be related by 
\begin{equation}
P_{\rm st,[J_{\rm st}]}({\bf x})= \hat P_{\rm st,[-J_{\rm st}]}({\bf x}).
\label{eq:Pf=Pb}
\end{equation}
That is, for an equilibrium state ($J_{st}=0$), the asymptotic forward and backward probabilities should be the same. For a more general out-of-equilibrium stationary state, the forward and backward probabilities are related by an inversion in the direction of the stationary current probability. The central question is which is the backward stochastic process that fulfills this requirement or, in other words, which is the drift $\hat {\bf f}({\bf x})$ that produces Eq.~(\ref{eq:Pf=Pb}). 
From Eqs.~(\ref{eq:Current}) and~(\ref{eq:CurrentBackward}), imposing ${\bf J}_{\rm st}=-\hat{\bf J}_{\rm st}$ and using Eq.~(\ref{eq:Pf=Pb}), we obtain 
\begin{equation}
\hat f_i({\bf x})= f_i({\bf x}) +\left(2\alpha-1\right)\; g_{k\ell}({\bf x})\partial_k g_{i\ell}({\bf x}).
\end{equation} 

Therefore, the time reversal transformation ${\cal T}$ which makes physical sense, meaning that it 
produces a backward evolution which  converges to a unique equilibrium distribution, is  given by  
\begin{equation}
{\cal T} = \left\{
\begin{array}{lcl}
{\bf x}(t) &\to & {\bf x}(-t)   \\   & & \\
\alpha &\to & (1-\alpha) \\ & & \\
f_i &\to&  f_i  +\left(2\alpha-1\right)\; g_{k\ell}\partial_k g_{i\ell}
\end{array}
\right.
\label{eq.T}
\end{equation}
Since $1-2\alpha$ is odd under the transformation $\alpha\to 1-\alpha$, the time reversal operator defined in this way satisfies 
${\cal T}^2=I$, as it should be. As we have previously stressed, in the
Stratonovich prescription $\hat f= f$  so, in this case,  the time
reversal operator ${\cal T}$ simply corresponds to change ${\bf x}(t)\to {\bf x}(-t)$. In 
any other prescription the definition of ${\cal T}$ is more involved, given by
Eq.~(\ref{eq.T}). 

Interestingly, the shift in the drift necessary to define the backward stochastic process is related with the difference in the stochastic trajectories of the forward and backward directions, computed in 
Eq.~(\ref{eq:DeltaXalpha}) and pictorially described in Figure~(\ref{fig:timereversal}). Indeed, we find that
\begin{equation}
\left[f_i({\bf x})-\hat f_i({\bf x})\right] \Delta t = \Delta_\alpha x_i(t),
\label{eq:deltaXdrift}
\end{equation}
where $\Delta_\alpha x_i(t)$ is given by Eq.~(\ref{eq:DeltaXalpha}).
Thus, the shift in the backward drift compensates the fact that the forward and backward stochastic trajectories do not end at the same point. 

%%%%%%%%%%%%%%%%%%%%%%%%%%%
\section{Generating functional}
\label{sec:GeneratingFunctional}
%%%%%%%%%%%%%%%%%%%%%%%%%%%
Alternatively  from Langevin and Fokker-Plank equations, we study in this section the path integral formalism. We generalize the techniques developed in Refs.~\cite{arenas2010,arenas2012,Arenas2012-2} for a single variable,  to the case of a $n$-dimensional vector stochastic variable satisfying a system of SDE given by Eq.~(\ref{eq:LangevSystem}).

We are interested in computing $n-\mbox{point}$ correlation functions \[\left\langle x_{i_1}(t_1) \dots x_{i_n}(t_n) \right\rangle, \mbox{      for } i_k \in [1,\ldots,n]. \]
To compute these correlation functions we should know the $n^{\rm th}$-order joint probability function of the random vector variable ${\bf x}(t)$. An alternative equivalent procedure is to solve the system of Langevin Eqs.~(\ref{eq:LangevSystem}) and compute
\begin{equation}
\left\langle x_{i_1}(t_1) \dots x_{i_n}(t_n) \right\rangle \equiv  \left\langle \bar x_{i_1[\eta]}(t_1) \dots \bar x_{i_n[\eta]}(t_n) \right\rangle_{\boldsymbol{\eta}},
\end{equation}
where ${\bf\bar x}_{[\eta]}(t)$ is a solution of (\ref{eq:LangevSystem}) for a particular realization of the noise and certain initial condition $\textbf{x}(t_0)=\textbf{x}_0$. 
\[<\ldots>_{\boldsymbol{\eta}} \; = \int \mathit{D\eta_1}\mathit{D\eta_2}\ldots\mathit{D\eta_n}\ \ldots \ e^{-\int dt\ \boldsymbol{\eta} \cdotp \boldsymbol{\eta}}\] means the stochastic mean value in the  variable $\boldsymbol{\eta}$.
Correlation functions can be obtained from a generating functional
\begin{equation}
 Z[{\bf J}(t)] =  \left\langle e^{\int dt\ {\bf J} \cdotp {\bf\bar{x}}_{[\eta]}(t)}\right\rangle_{\boldsymbol{\eta}}
\label{ZJxbar}
\end{equation}
by simply differentiating with respect to the source ${\bf J}(t)$,
\begin{equation}
 \left\langle \bar x_{i_1[\eta]}(t_1) \dots \bar x_{i_2[\eta]}(t_n) \right\rangle_{\boldsymbol{\eta}}  = \frac{\delta ^n Z[{\bf J}]}{\delta J_{i_n}(t_n) \dots \delta J_{i_1}(t_1) } \Bigg \vert_{{\bf J}=0} \ .
\end{equation}

Our goal is to find a functional representation for $Z({\bf J})$, avoiding the problem of explicitly solving the system of Langevin equations.  Also, we want to present a completely functional formalism, obtained without performing any discretization of the SDE. In this way, we don't need to deal explicitly with Wiener integrals. For this purpose, we begin by introducing  a functional integral over ${\bf x}(t)$ and a delta-functional which constraints its value to a solution of the SDE. Thus, we can rewrite Eq.~(\ref{ZJxbar}) in the following form,
\begin{equation}
 Z[{\bf J}] = {\left\langle \int [D{\bf x}]\; \delta^n[{\bf x}(t)- {\bf \bar{x}}_{[{\bf\eta}]}(t)] \ e^{\int dt {\bf J}(t)
 \cdot {\bf x}(t) }\right\rangle }_{\boldsymbol{\eta}}
\end{equation}
or, since the only noisy terms are contained in  ${\bf \bar x}_{[\eta]}(t)$,
\begin{equation}
 Z[{\bf J}] =  \int [D{\bf x}] e^{\int dt {\bf J}(t)
 \cdot {\bf x}(t) }\; \left\langle\delta^n[{\bf x}(t)- {\bf \bar{x}}_{[{\bf\eta}]}(t)]\right\rangle_{\boldsymbol{\eta}} \ .
 \label{eq:Zdeltaxeta}
\end{equation}
The advantage of this expression is that the source ${\bf J}$ is no longer coupled with the solution of 
the SDE. The next step is to eliminate the explicit dependence on this solution by using the following property of the delta-functional,
\begin{equation}
\delta^n[{\bf x}(t)- {\bf \bar{x}}_{[{\bf\eta}]}(t)]=  \delta^n[{\bf \hat{O}}({\bf x})]\;{\rm det}\left(\frac{\delta \hat{O}_i}{\delta x_k}\right),\; \;k=1,\ldots,n,
\label{delta-property}
\end{equation}
where ${\bf \bar{x}}_{[\eta]}(t)$ is a root of ${\bf \hat{O}}({\bf x})$, \emph{i.e.}, 
${\bf \hat{O}}({\bf \bar{x}}_{[\eta]}) = 0$.
We assume here that the operator $\hat {\bf O}$ has only one root. So, we are supposing that, for a particular realization of the noise, the trajectory is completely specified by the initial conditions ${\bf x}(t_0)={\bf x}_0$.
The form of operator ${\bf \hat{O}}({\bf x})$ is not uniquely determined. A simple possible choice is 
\begin{equation}
 \hat{O}_i({\bf x}) = \frac{dx_i(t)}{dt} - f_i({\bf x}) - g_{ij}({\bf x})\eta_j (t).
 \label{O}
\end{equation}
Correspondingly, the differential operator $\delta \hat{O}_i/\delta x_k$ is given by
\begin{equation}
\frac{\delta \hat{O}_i({\bf x}(t))}{\delta x_k(t')} = \left[ \delta_{ik}\frac{d}{dt} - \partial_k f_i({\bf x}) - \partial_k g_{ij}({\bf x})\eta_j (t)\right] \delta (t-t'),
\label{O'}
\end{equation}
where $\partial_k$ means, as usual,  partial differentiation with respect to $x_k$.
Substituting Eq.~(\ref{delta-property}) into Eq.~(\ref{eq:Zdeltaxeta}), the generating functional now reads, 
\begin{equation}
 Z[{\bf J}] =  \int [D{\bf x}] e^{\int dt {\bf J}(t)
 \cdot {\bf x}(t) }\; \left\langle\delta^n[{\bf \hat{O}}] \det\left(\frac{\delta \hat{O}_i}{\delta x_k}  \right)\right\rangle_{\boldsymbol{\eta}}  \ , 
\label{eq:Z2}
\end{equation}
with the definitions of the Eqs.~(\ref{O}) and~(\ref{O'}). At this point, we have completely eliminated from the generating functional any reference to the explicit solution of the SDE. 

The main step is to correctly obtain the statistical mean value over the noise $\boldsymbol{\eta}$.  In the multiplicative case, not only the delta-functional but also the determinant are $\boldsymbol{\eta}$--dependent and the mean value in Eq.~\eqref{eq:Z2} can not be directly performed. As a consequence, for the aforementioned purpose, we will represent the delta-functional and the determinant through integral expressions using auxiliary variables.  
Introducing a vector auxiliary variable $\boldsymbol{\varphi}(t)=(\varphi_1(t), \ldots,\varphi_n(t) )$ we can represent the delta-functional as a Fourier functional integral in the following form, 
\begin{equation}
\delta\left[{\bf \hat O}({\bf x}) \right] = \int [D \boldsymbol{\varphi}] \;  
 e^{-i\int dt \  \varphi_i\left[ \dot x_i(t) - f_i - g_{ij}\eta_j\right]}.
\label{delta}
\end{equation}
where the ``dot'' means time differentiation.

The representation of the determinant is more involved. In the same way that inverse determinants can be represented by Gaussian integrals, a determinant itself of any matrix can be represented as a Gaussian integral over anti-commuting variables (see appendix~\ref{sec:Determinants}). 
Thus, we introduce a vectorial  function $ \boldsymbol{\xi}(t) = (\xi_1(t),\ldots,\xi_n(t))$ and its conjugate $\bar{\boldsymbol{\xi}}(t) = (\bar \xi_1(t),\ldots,\bar \xi_n(t))$, which obey the Grassmann algebra
\begin{equation}
\{\xi_i(t),\xi_j(t')\}=\{\bar\xi_i(t),\bar\xi_j(t')\}=\{\xi_i(t),\bar\xi_j(t')\}=0,
\end{equation}
where $\{~, \}$ represents anti-commutators. These relations imply, in particular, $\xi_i(t)^2=\bar\xi_i(t)^2=0$. In terms of these auxiliary variables, 
the determinant can be represented as (for mathematical details see Ref.~\cite{Zinn-Justin} and appendix~\ref{sec:Determinants})
\begin{eqnarray}
\lefteqn{
\det \left(\frac{\delta \hat{O}_i({\bf x})}{\delta x_k} \right) = } \nonumber \\ 
&=&\int [D \boldsymbol{\xi}] [D \bar{\boldsymbol{\xi}}]
   \ e^{\int dt dt'\;\bar{\xi}_i(t) \left(\frac{\delta \hat{O}_i({\bf x}(t))}{\delta x_k(t')} \right)\xi_j(t') } \nonumber  \\
 &=&  \int  [D \boldsymbol{\xi}] [D \bar{\boldsymbol{\xi}}]  \ e^{\int dt \bar{\xi}_i\dot\xi_i-\int dt \bar{\xi}_i \left[ \partial_k f_i - \partial_k g_{ij}\eta_j\right] \xi_k} .
\label{detgrassman}
\end{eqnarray}
Substituting representations~(\ref{delta}) and~(\ref{detgrassman})  into Eq.~(\ref{delta-property}), we find
\begin{eqnarray}
\lefteqn{
\left\langle \delta [{\bf x} - {\bf \bar{x}}_{{[\eta]}}]\right\rangle _{\boldsymbol{\eta}} =
\int [D \boldsymbol{\xi}] [D \bar{\boldsymbol{\xi}}] [D \boldsymbol{\varphi}]\;  \times }\nonumber \\
&\times& e^{\int dt\;\left\{ \bar{\xi}_i(t) \dot\xi_i(t) -\bar{\xi}_i(t)\partial_k \xi_k(t) f'(x) -i  \varphi_i(t) \left[\dot x_i(t) - f_i(x) \right]\right\}}
\nonumber \\
&\times& \left\langle  e^{-\int dt\;\left[ \bar{\xi}_i(t)\partial_k g_{ij} \xi_k(t) -i \varphi_i(t) g_{ij}(x) \right]\eta_j(t)}\right\rangle _{\boldsymbol{\eta}} \ .
\label{delta-exvalue}
\end{eqnarray}

Since the noise distribution is Gaussian, it is now immediate to compute the average over the noise and we find,
\begin{eqnarray}
\lefteqn{ \left\langle  e^{-\int dt\;\left[ \bar{\xi}_i \partial_k g_{ij}\eta_j \xi_k -i \varphi_i g_{ij}\eta_j\right]}\right\rangle _{\boldsymbol{\eta}} } \nonumber \\
& \sim & \ \exp \left\{ \frac{1}{2} \int dt \left( \partial_m g_{lj} \partial_k g_{ij} \bar{\xi}_l \xi_m \bar{\xi}_i \xi_k -
\right.\right. \nonumber \\
&&÷÷÷÷÷÷÷- \left.\left. 2i\varphi_l g_{lj} \partial_k g_{ij} \bar{\xi}_i \xi_k - \varphi_i g_{ij} \varphi_l g_{lj}\right) \right\} .
\label{average}
\end{eqnarray}

Introducing Eq.~(\ref{delta-exvalue}) into Eq.~(\ref{eq:Z2}) and using Eq.~(\ref{average}), we finally arrive at the desired representation for the generating functional of correlation functions,
\begin{equation}
 \mathit{Z}[{\bf J}] = \int [Dx] [D \boldsymbol{\varphi}] [D \boldsymbol{\xi}] D\bar{\boldsymbol{\xi}}] \;\; e^{-S + \int dt {\bf J}\cdot  {\bf x}},
\label{eq:funcionalgenerador}
\end{equation}
where the ``action'' $S$ is given by

\begin{widetext}
\begin{eqnarray}
S & = &  \int  dt \left\{  i\varphi_l(t) \left[ \dot{x}_l(t) - f_l(x)  + g_{lj}(x)\partial_k g_{ij}(x)\bar{\xi}_i(t)\xi_k(t) \right]  + \frac{1}{2}\varphi_i(t) \varphi_l(t) g_{ij}(x) g_{lj}(x)  \right. \nonumber \\
& - &  \left.\bar{\xi}_i(t) \dot{\xi}_i(t) + \partial_k f_i(x) \bar{\xi}_i(t) \xi_k(t)-  \frac{1}{2}\partial_m g_{lj}(x)\partial_k g_{ij}(x)\bar{\xi}_l(t)\xi_m(t) \bar{\xi}_i(t)\xi_k(t) \right\} .
\label{action}
\end{eqnarray}
\end{widetext}

This expression is the major result of this section. It states that we can represent a stochastic multiplicative process as a functional integral over a set of commuting and anti-commuting variables. The multiplicative character of the process can be visualized through terms proportional to partial derivatives of the diffusion matrix, 
$\partial_k g_{ij}$. In the first line of Eq.~(\ref{action}) this term couples the response variables $\varphi_i$ with the Grassmann variables. This a  typical signature of multiplicative processes. 
The other term, in the second line of Eq.~(\ref{action}), is an interaction (quartic) term for the Grassmann variables. This term is only possible in a vectorial case. For a single stochastic variable, it is automatically zero due to the property $\xi(t)^2=\bar \xi(t)^2=0$. 

The principal advantage of this representation is that it does not depend on any particular stochastic prescription. In some sense, the complex stochastic calculus described in appendix~\ref{sec:alphacalculus} is now codified in the Grassmann algebra.  One of its consequences is that prescription-dependent transformations, such as time reversal, can be written in a simpler form in this representation.  In fact, the time reversal transformation, Eq.~(\ref{eq.T}), has a linear form in terms of Grassmann variables,
\begin{equation}
{\cal T} = \left\{
\begin{array}{lcl}
x_i(t) &\to  &  x_i(-t)  \\ 
& & \\
\varphi_i(t) & \to  & \varphi_i(-t) -2i \left(g^2\right)^{-1}_{ij}\; \dot x_j(-t) \\
& & \\
\xi_i(t) &\to & \bar\xi_i(-t)   \\
& & \\
\bar\xi_i(t) &\to &- \xi_i(-t)
\end{array}
\right.
\label{eq.TG}
\end{equation}  

Of course, since for each value $\alpha$ of the stochastic prescription we have a different stochastic process, the action in Eq.~(\ref{action}) represents a family of stochastic processes and not just one of them. To better see it, in the next section we perform the integration over the Grassmann variables. 

%%%%%%%%%%%%%%%%
\section{Grassmann integration, Green's functions and prescription dependent representation}
\label{sec:GrassmanIntegration}
%%%%%%%%%%%%%%%
 
In the preceding formalism we did not need to specify in which stochastic prescription we were working. In fact, the action (\ref{action}) is independent of any stochastic prescription. 
In this way, we do not need to bother with It\^o calculus when dealing with this functional formalism. In some way, all the difficulties of stochastic calculus are codified in the Grassmann algebra. Of course, the problems associated with the choice of a prescription should reappear if we eliminate the Grassmann variables from the theory. However, since we did not discretized the time domain, the ambiguities related with the $\alpha$-prescription will appear in a different form.  To see this, let us integrate out the set of Grassmann variables to obtain an action depending only on the commuting variables ${\bf x}$ and $\boldsymbol{\varphi}$. 
 
Let us write the Grassmann sector of the functional integral, 
\begin{eqnarray}
I_{\rm G}
& = & \int [D\bar{\boldsymbol{\xi}}] [D \boldsymbol{\xi}] \; e^{\int dt \bar{\xi}_i \dot{\xi}_i} e^{-\int dt \left[\partial_k f_i - i\varphi_l g_{lj} \partial_k g_{ij} \right] \bar{\xi}_i \xi_k} \times \nonumber \\
&\times&  e^{ \int dt \frac{1}{2}\partial_m g_{lj}\partial_k g_{ij}\bar{\xi}_l \xi_m \bar{\xi}_i\xi_k}
\label{GrassmIntegral}
\end{eqnarray}

Just to illustrate the method, let us consider only the first line of Eq.~(\ref{GrassmIntegral}). We will return to the quartic term later.  By performing a Taylor expansion of the second exponential we have
\begin{eqnarray}
I_{\rm G}^0& = & \int [D\bar{\boldsymbol{\xi}}] [D \boldsymbol{\xi}] \ e^{\int dt \bar{\xi}_i \dot{\xi}_i} \left\{ 1 - \int dt \mathit{F^0_{ik}}(t) \bar{\xi}_i(t) \xi_k(t)  \right.  \nonumber \\
& & \left. + \ \frac{1}{2} \int dt dt' \mathit{F^0_{ik}}(t)
\mathit{F^0_{\ell m}}(t') \bar{\xi}_i(t) \xi_k(t) \bar{\xi}_\ell(t') \xi_m(t')  \right.  \nonumber \\
& & \left. - \ \ldots \right\} \nonumber \\
& = & N \left\{ 1 - \int dt \mathit{F^0_{ik}}(t) \left\langle \bar{\xi}_i(t) \xi_k(t) \right\rangle  \right.  \nonumber \\
& & \left. + \ \frac{1}{2} \int dt dt' \mathit{F^0_{ik}}(t)
\mathit{F^0_{\ell m}}(t') \left\langle \bar{\xi}_i(t) \xi_k(t)
\bar{\xi}_\ell(t') \xi_m(t') \right\rangle   \right. \nonumber\\
& & \left. - \ \ldots \right\},
\label{Taylor1}
\end{eqnarray}
where $\mathit{F^0_{ik}}(t) = \partial_k f_i(t) - i\varphi_l(t) g_{lj}(t) \partial_k g_{ij}(t)$, $N=\det(d/dt)$
and the correlation functions $\langle\dots\rangle$ are computed using the Gaussian weight 
$\exp(\int dt \bar{\xi}_i \dot{\xi}_i)$.
We use Wick's theorem to factorize the multi-point correlation functions in terms of two-point correlations, 
\begin{eqnarray}
\lefteqn{
\left\langle \bar{\xi}_i(t) \xi_k(t) \bar{\xi}_\ell(t') \xi_m(t')
\right\rangle  =  \left\langle \bar{\xi}_i(t) \xi_k(t)
\right\rangle \left\langle \bar{\xi}_\ell(t') \xi_m(t') \right\rangle + }\nonumber \\
&+&\overbrace{\left\langle \bar{\xi}_i(t)
\bar{\xi}_\ell(t')\right\rangle}^0
\ \overbrace{\left\langle \xi_k(t)\xi_m(t') \right\rangle}^0 +\left\langle \bar{\xi}_i(t)
\xi_m(t') \right\rangle \left\langle \xi_k(t) \bar{\xi}_\ell(t')
\right\rangle ,   \nonumber \\
& & 
\end{eqnarray}
and we finally have
\begin{eqnarray}
I_{\rm G}^0
&=& N \left\{ 1 - \int dt \mathit{F^0_{ik}}(t) \left\langle \bar{\xi}_i(t) \xi_k(t) \right\rangle \right.  \nonumber \\
& & \left. + \ \frac{1}{2} \int dt dt' \mathit{F^0_{ik}}(t)
\mathit{F^0_{\ell m}}(t') \left\langle \bar{\xi}_i(t) \xi_k(t) \right\rangle \left\langle \bar{\xi}_\ell(t') \xi_m(t') \right\rangle \right.   \nonumber \\
& & \left. - \ \frac{1}{2} \int dt dt' \mathit{F^0_{ik}}(t)
\mathit{F^0_{\ell m}}(t') \left\langle \bar{\xi}_i(t) \xi_\ell(t') \right\rangle
\left\langle \bar{\xi}_m(t') \xi_k(t) \right\rangle
\right. \nonumber \\
& &  \left. - \ \ldots \right\}.
\label{Taylor-Wick}
\end{eqnarray}
Considering that the dynamical part of the Grassmann action is Gaussian, we have that the two-point correlation function is simply given by 
\begin{equation}
\left\langle \bar{\xi}_i(t) \xi_k(t')\right\rangle =  \delta_{ik}  G(t,t'),
\end{equation}
where $G(t,t')$ is the Green's function of the first derivative operator $d/dt$, \emph{i.e.}, 
\begin{equation}
\frac{d G(t,t')}{dt}=\delta(t-t').
\end{equation}

As usual, we need a prescription to compute the Green's function, generally given by initial conditions. For causality reasons we decided to work with the retarded Green's function, for which $G_{\rm R}(t,t')=0$ if $t<t'$. In such a case, 
\begin{equation}
G_{\rm R}(t,t')=\Theta(t-t'),
\label{eq:GR}
\end{equation}
where $\Theta(t)$ is the Heaviside distribution.

Going back to expression~(\ref{Taylor-Wick}) it is simple to realize that the last term (and the subsequent ones with the same type of two-point correlations) is zero because 
\begin{equation}
G_{\rm R}(t,t')G_{\rm R}(t',t)=0,
\label{GRGR}
\end{equation}
except in the null measure set $t=t'$. We can now re-exponentiate the remaining part of the series and we obtain
\begin{eqnarray}
I_{\rm G}^0 = N e^{-G_{R}(0)\int dt \left[\partial_i f_i - i\varphi_l g_{lj}
\partial_i g_{ij} \right] }.
\label{GrassmIntegral_1}
\end{eqnarray}

In a very similar way we can compute the integral Eq.~(\ref{GrassmIntegral}), considering both the quadratic and the quartic terms. Due to the Gaussian properties of the averages and using the retarded Green's function,  it is simple to show
\begin{eqnarray}
I_{\rm G}&=& \langle e^{\int dt  - \left[\partial_k f_i - i\varphi_l g_{lj} \partial_k g_{ij} \right] \bar{\xi}_i \xi_k}\rangle \langle e^{ \frac{1}{2}\int dt\partial_m g_{lj}\partial_k g_{ij}\bar{\xi}_l\xi_m \bar{\xi}_i\xi_k   }\rangle \nonumber \\
& = & I^0_{\rm G}\times I_{\rm G}^1 ,
\label{GrassmIntegral_2}
\end{eqnarray}
with
\begin{equation}
I_{\rm G}^1 = \int [D \boldsymbol{\bar{\xi}}] [D \boldsymbol{\xi}] \ e^{\int dt
\bar{\xi}_i \dot{\xi}_i} e^{\frac{1}{2} \int dt F^1_{lmik}(t) \bar{\xi_l}
\xi_m \bar{\xi_i} \xi_k} \label{I1} ,
\end{equation}
where $F^1_{lmik}(t) = \partial_m g_{lj} \partial_k g_{ij}$.
Expanding the exponential in powers of $F^1_{lmik}(t)$, using the Wick's theorem and taking into account 
the condition~(\ref{GRGR}) for the retarded Green's function, we find  
\begin{equation}
I^1_{\rm G} = N  \exp\left\{\frac{1}{2} \ G_{\rm R}^2(0) \int dt  F^1_{iklm}(t) 
\left( \delta_{lm} \delta_{ik} - \delta_{lk} \delta_{im} \right) \right\}.
\label{calculo_I1}
\end{equation}

\begin{figure}
\includegraphics[height= 4 cm]{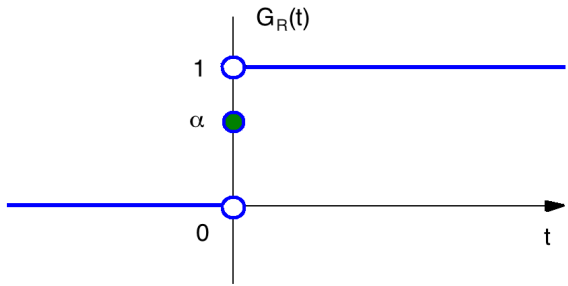}
\caption{Retarded Green's function}
\label{fig.GR}
\end{figure}

We observe that the integration over Grassmann variables is well defined up to an undetermined constant 
given by $G_{\rm R}(0)$. As the Green's function is discontinuous at the origin, it is not well defined at this location and we need a prescription to do that.  In Figure~(\ref{fig.GR}) we show the retarded Green's function with a proper definition at the origin $G_{\rm R}(0)=\alpha$, with $0\le\alpha\le 1$.
With this definition, we finally obtain,
\begin{widetext}
\begin{eqnarray}
S & = & \int dt \left\{ \alpha \partial_k f_k(x) + \frac{1}{2}\varphi_i(t) \varphi_l(t) g_{ij}(x) g_{lj}(x) + i\varphi_l(t) \left[ \partial_t x_l - f_l + \alpha g_{lj}(x)\partial_i g_{ij}(x) \right] \right.  \nonumber\\
&  & \left.  + \frac{1}{2} \alpha^2 \left[ \partial_m g_{kj}(x) \partial_k g_{mj}(x) - \partial_m g_{mj}(x) \partial_i g_{ij}(x) \right]  \right\} \ .
\label{eq:action-alpha}
\end{eqnarray}
\end{widetext}
This result coincides with the one computed by direct discretization of the system of Langevin equations~\cite{Lubensky2007}. This fact allows us to identify the definition of the Green's function at the origin with 
the Generalized Stratonovich prescription used to define Wiener integrals. 

 \begin{figure}
\includegraphics[height= 4 cm]{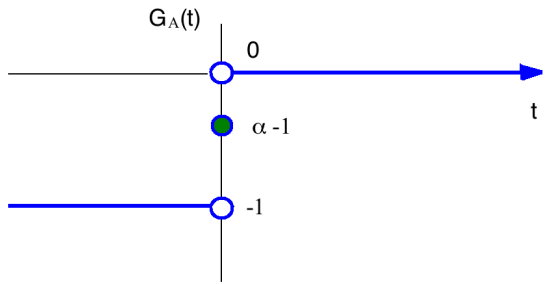}
\caption{Advanced Green's function}
\label{fig.GA}
\end{figure}

This action, written only in terms of the commuting variables $\{ {\bf x}(t),\boldsymbol{\varphi}(t) \}$, depends now on the stochastic prescription $\alpha$. In order to operate with it we need to take into account the stochastic calculus described in appendix~\ref{sec:alphacalculus}, summarized in the generalized chain-rule, Eq.~(\ref{CR}).
At the same time, to study time reversal trajectories in this representation, we need to consider the transformation in 
Eq.~(\ref{eq.T}). 

It is instructive to observe how the particular transformation under time reversal 
$\alpha\to 1-\alpha$  emerges in the language of Grassmann variables.  The parameter $\alpha$ appears as the definition at the origin of the {\rm retarded} Green's function of the Grassmann variables. However, a time reversal transformation transforms the {\rm retarded} Green's function into the {\em advanced} one, 
$G_R(t)\to G_A(t)$. As is known, different Green's functions of the same operator can only differ in a solution of the homogeneous equation. 
In our case, the differential operator for Grassmann variables is simply the first-order total time derivative $d/dt$. Therefore, the general solution of the homogeneous equation is just a constant. 

The retarded Green's function is $\Theta(t)$ as showed in Eq.~(\ref{eq:GR}). 
On the other hand, the advanced Green's function is $G_A(t)=-\Theta(-t)$. Then, for $t\neq 0$, we trivially have, 
\begin{equation}
G_R(t)-G_A(t)=\Theta(t)+\Theta(-t)=1.
\end{equation}
If we define at the origin $G_R(0)=\alpha$, as shown in figure~(\ref{fig.GR}), then, we must have $G_A(0)=\alpha-1$, as shown in figure~(\ref{fig.GA}). This implies for the definition of the Heaviside functions, $\Theta(0^+)=\alpha$ and $\Theta(0^-)=1-\alpha$.
In this way, we recover the transformation $\alpha\to 1-\alpha$ under time reversal, deduced in section~\ref{sec:Langevin} by a careful definition of the Wiener integral.
Just for consistency, let us analyze the behavior of a semi-sum of the retarded and the advanced Green's functions. By definition, the result should be another Green's function. 
In fact, we have
\begin{eqnarray}
G(t)=\frac{G_R(t)+G_A(t)}{2} 
&=&  \frac{1}{2}\left(\Theta(t)-\Theta(-t)\right) \nonumber \\
&=&\frac{1}{2} \ {\rm Sign}(t),
\end{eqnarray}
where $dG/dt=\delta(t)$ and $G(t)=-G(-t)$ for $t\neq 0$.
With our definitions of the Heaviside functions at the origin, we obtain
\begin{equation}
G(0)= \frac{2\alpha-1}{2}.
\end{equation}
This expression is odd under time reversal $\alpha\to 1-\alpha$, as it should be. 

%%%%%%%%%%%%%%%%%%%%%%
\section{Examples and applications}
\label{sec:Examples}
%%%%%%%%%%%%%%%%%%%%%%

In this section, we present three specific applications with the aim of showing that the functional formalism developed in this paper is quite general and can be directly used to describe several kinds of stochastic processes. 

%%%%%%%%%%%%%%%%%%%%%%%%%
\subsection{Brownian particle near a wall}

Let us begin by the simplest case of a single stochastic variable submitted to external conservative forces and a noisy environment in the presence of boundary conditions. 
Consider a colloidal particle  immersed in water and diffusing in a closed sample cell above a planar wall~\cite{Volpe2010}.  The wall is perpendicular to the $z-$axis and it is  placed at 
$z=0$.  We are interested in the dynamics of  $z(t)$ that measures the distance between the particle and the wall. 

The system is usually modeled by an overdamped Langevin equation of the form  
\begin{equation}
dz=\frac{F(z)}{\gamma(z)} dt+\sqrt{2D_{\perp}(z)} dW.
\label{eq:colloidal-Langevin}
\end{equation}
$F(z)$ is an external conservative force which has two contributions,
\begin{equation}
F(z)= B e^{-\kappa z}-G_{\rm eff}\;.
\end{equation}
The first term parametrizes the  electrostatic component that decays exponentially away from the wall within a typical distance $\kappa^{-1}$. The prefactor $B$ depends on the surface charge densities. The second term represents the effective constant gravitational force. This conservative force can be obtained from a classical potential energy,
\begin{equation}
V(z)=\frac{B}{\kappa} e^{-\kappa z}+ G_{\rm eff} z \, ,
\end{equation}
which has a global  minimum for  $z > 0$, provided $B > G_{\rm eff}$. This minimum determines the mean distance between the wall and the particle,  around which we study fluctuations.
The friction coefficient $\gamma(z)$ and the diffusion function $D_{\perp}(z)$ are related by the Einstein relation 
$\gamma(z) D_{\perp}(z)=1/\beta$, where 
$\beta=1/k_B T$ is, essentially, the inverse temperature of the thermal bath. Finally, $dW$ represents a usual Wiener process.   
In the absence of a wall, or  far away from $z=0$, the Brownian motion is characterized by a diffusion constant 
$D_{\infty}$. However, near the wall, fluctuations are strongly suppressed, leading to  a state--dependent diffusion process. The function $D_{\perp}(z)$ was analytically  computed~\cite{Brenner1961} using Navier-Stokes equations and was also measured~\cite{Volpe2010} in colloidal particles with an excellent agreement. 
This function can be successfully parametrized by 
\begin{equation}
D_{\perp}(z)\sim \frac{D_{\infty}}{1+\frac{9}{8} (a/z)}\; ,
\end{equation}
where $a$ is the radius of the colloidal particle. 

The overdamped Langevin equation~(\ref{eq:colloidal-Langevin}) has the same form of Eq.~(\ref{eq:LangevSystem}) with $n=1$, 
$f(z)= F(z)/\gamma(z)$ and the diffusion matrix with a single component,  $g(z)=\sqrt{2D_{\perp}(z)}$.   
It is interesting to compute the quantity 
\begin{equation}
g(z) g'(z)=D'_{\perp}(z)\sim \frac{8}{9}\frac{D_\infty}{a}\left[ \frac{1}{1+\frac{8}{9}(z/a)}\right]^2  ,
\label{eq:ggprime}
\end{equation}
that enters in the expression~(\ref{eq:Current}) for the probability current. In Eq.~(\ref{eq:ggprime}) and in the following we use the notation $g'$ to indicate total derivative with respect to $z$. We see that, for large values of $z$, 
$g(z)g'(z)$ tends to zero since, far away from the wall, the noise is essentially additive. However, very near the wall, 
$g(z)g'(z)$ approaches a constant, signaling a truly multiplicative noise situation. Interestingly, in this case we can exactly solve the static limit of the  Fokker-Plank equation~(\ref{eq:FP}), obtaining the equilibrium probability distribution 
$P_{\rm eq}(z)= N \exp(-\beta U_{\rm eq})$, where $N$ is a normalization constant and the equilibrium potential is 
given by~\cite{arenas2012,Arenas2012-2}
\begin{equation}
U_{\rm eq}(z)= V(z)+ (1-\alpha)k_B T \ln(2D_{\perp}(z))\ .
\label{eq:Ueq}
\end{equation}
It can be seen, as expected, that the equilibrium solution depends on $\alpha$. Each value of $\alpha$ represents a different dynamical evolution. This is a hallmark of a multiplicative process. From  Eq.~(\ref{eq:Ueq}), it is evident that the Boltzmann equilibrium distribution $P_{\rm eq}\sim \exp(-\beta V)$ is only reached for $\alpha=1$. 
This fact was experimentally observed in Ref.~\cite{Volpe2010}.  Thus, in order to correctly model the dynamics of a colloidal particle near a wall we need to consider the stochastic differential equation~(\ref{eq:colloidal-Langevin}) in the H\"anggi-Klimontovich interpretation $\alpha=1$. Of course, if for any reason we would prefer to work in any other stochastic prescription, we should add to Eq.~(\ref{eq:colloidal-Langevin})  a ``spurious drift term" proportional to  $D'_{\perp}(z)$,  using  Eq.~(\ref{eq:Translator}). 

One of the advantages of the functional formalism presented in this paper, is that we can work without specifying the prescription until the end of the calculation. This allows to use usual calculus rules provided we deal with Grassmann variables. 
It is simple to write the effective action by  considering Eq.~(\ref{action}) for just one variable.  The result  in the extended functional space $\{z(t), \varphi(t), \bar\xi(t), \xi(t)\}$  takes the form
\begin{eqnarray}
S & = &  \int  dt \left\{ - \bar{\xi}\dot{\xi} +  f' \bar{\xi} \xi   \right. \nonumber \\ 
& +& \left. i\varphi \left[ \dot{z} - f  + gg'\bar{\xi}\xi \right]  + \frac{1}{2}\varphi^2 g^2  
\right\} ,
\label{eq:action1d-Grassman}
\end{eqnarray}
where the ``dot'' means total time differentiation and the ``prime'', $z$-differentiation. 
The  multiplicative character of the noise is evident in the coupling between the Grassmann variables $\bar\xi\xi$ and the response variable $\varphi$,  as can be observed  in the second line of Eq.~(\ref{eq:action1d-Grassman}).  The term with four Grassmann variables  is absent in the single variable case, due to the property $\xi^2=\bar\xi^2=0$.
Thus, with this action it is possible to compute correlation functions by means of perturbative as well as non-perturbative techniques, without any reference to non-trivial stochastic calculus. The anticommuting properties of the Grassmann variables automatically take into account the subtleties of the stochastic calculus.  Alternatively,   we can integrate out the Grassmann variables,  as well as the response variable,  to work with a functional formalism just in terms of  the variable $z(t)$.  Making the integrations,  fixing $\langle\bar\xi(t)\xi(t)\rangle=G_R(0)=1$ and  replacing the values of $f(z)$ and $g(z)$, 
 we obtain the effective action 
 \begin{equation}
S =\int dt \left\{\frac{ \left(\dot z - \beta D_{\perp} F + D'_{\perp}\right)^2}{2D_{\perp}}
+  \beta ( D_{\perp} F)' \right\} \ .
\label{eq:action1d-alpha}
\end{equation}
This equation represents the probabilistic weight attached to each stochastic trajectory.  Although Eq.~(\ref{eq:action1d-alpha}) seems to be simpler than Eq.~(\ref{eq:action1d-Grassman}), it should be taken into account that, for any manipulation of the action~(\ref{eq:action1d-alpha}), we need to use the generalized stochastic calculus described in the Appendix~\ref{sec:alphacalculus}. 

%%%%%%%%%%%%%%%%%
\subsection{Non-Markov processes driven by higher derivatives Langevin equations}

In this example we consider a single-variable non-Markov process, \emph{i.e.}, a process whose dynamics is influenced by  memory effects. For the sake of simplicity, we consider a system driven by a single Langevin equation with second-order derivatives. As is well known,  in this case the process is univocally defined, the It\^o -- Stratonovich dilemma no longer applies~\cite{vanKampen, vanKampenIto, Fox1978}. 
However, we would like to show how this comes about in the frame of our functional formalism in terms of Grassmann variables.

Let us consider a single stochastic variable $x(t)$, which satisfies the following SDE 
\begin{equation}
\frac{d^2x(t)}{dt^2}+\gamma \frac{dx(t)}{dt} -f(x)=g(x)\eta(t),
\label{NonMarkSDE}
\end{equation}
with $\langle\eta\rangle=0$ and $\langle\eta(t)\eta(t')\rangle=\delta(t-t')$.
As usual, $f(x)$ is a drift force, $g(x)$ defines the multiplicative character of the noise and  $\gamma$ is a friction coefficient.  We can transform this second-order equation into a couple of 
first-order equations by introducing new variables $x_1(t)=x(t)$ and $x_2(t)=dx_1/dt$, in such a way that 
\begin{eqnarray}
\frac{dx_1(t)}{dt}-x_2(t) & = & 0 ,\\
\frac{dx_2(t)}{dt}+\gamma x_2(t) & = & f(x_1) + g(x_1)\eta(t).
\end{eqnarray}
Comparing with Eq.~(\ref{eq:LangevSystem}) it is immediate to identify the drift vector
\begin{eqnarray}
f_1(x_1,x_2)&=&x_2, \\
f_2(x_1,x_2)&=&-\gamma x_2+f(x_1) 
\end{eqnarray}
and the diffusion matrix
\begin{equation}
\hat g({\bf x})=\left(
\begin{array}{cc}
0 & 0  \\
0 & g(x_1) 
\end{array}
\right).
\end{equation}
The structure of the diffusion matrix implies
\begin{equation}
g_{k\ell}\partial_k g_{i\ell}=0,
\end{equation}
and according with Eq.~(\ref{eq:gdg}) the process is $\alpha$ independent.
Another way to see this point is through the Fokker-Planck  Eqs.~(\ref{eq:FP}) and~(\ref{eq:Current}) associated with this process, which does not depend on $\alpha$.  Thus, although this is a multiplicative process, it is uniquely defined and it does not depend on the election of a particular prescription to define Wiener integrals. In addition, the time reversal process is characterized by the simple transformation ${\bf x}(t)\to {\bf x}(-t)$ because the drift stays the same  under transformation~\eqref{eq.T}. Consequently, from Eq.~\eqref{eq:deltaXdrift}, $\Delta_\alpha x_i(t) = 0$, indicating that the forward and backward time evolutions do have the same endpoints. The reason behind these facts is that it is not a Markov process, the second derivative induces some kind of memory in the system. 

Let us show how this fact is encoded in the structure of the generating functional, Eq.~(\ref{eq:funcionalgenerador}), with the action~(\ref{action}).  
 
Firstly, we  consider the term which couples the response variables $\varphi_i$ to Grassmann variables, 
\begin{equation}
i\varphi_\ell g_{k\ell} \partial_k g_{i\ell} \bar\xi_i\xi_k= i\varphi_2 g(x_1)\partial_1g(x_1)  \bar\xi_2\xi_1 .
\end{equation}
So, in this case, only $\varphi_2$ contributes to this term and it is proportional to a cross product of  Grassmann variables  
$\bar\xi_2\xi_1$. 
On the other hand, the quartic term in the  Grasmann variables is automatically zero since
\begin{equation}
\partial_{m}g_{lj}\partial_{k}g_{ij} \bar\xi_\ell\xi_m\bar\xi_i\xi_k=
(\partial_1 g)^2 \bar\xi_2\xi_1\bar\xi_2\xi_1\equiv 0
\end{equation}
and $\bar\xi^2=\xi^2=0$.
The rest of the quadratic terms in Grassmann variables simplify due to 
$\partial_1f_1=0$, $\partial_2f_2=-\gamma$, $\partial_1f_2=\partial_1 f$ and $\partial_2f_1=1$. 
Putting all these results together, integrating over the Grassmann variables and using the results
$\langle\bar\xi_1\xi_2\rangle=\langle\bar\xi_2\xi_1\rangle=0$ and $\langle\bar\xi_2\xi_2\rangle=\alpha$, we find that the only $\alpha$
 dependence is a constant $\exp(-\gamma\alpha)$ that can be absorbed in the normalization constant of the generating functional. Then, we end up with 
\begin{eqnarray}
S&=&\int dt \left\{ i\varphi_2[\dot x_2+\gamma x_2 -f(x_1)] \right. \nonumber \\
& & \left. +\ \frac{1}{2}g(x_1)^2\varphi_2^2 + i\varphi_1[\dot x_1-x_2] \right\}.
\end{eqnarray} 
Since there is no term proportional to $\varphi_1^2$, the last term in the second line of the preceding equation simply enforces the constraint 
$\delta(\dot x_1-x_2)$. Functionally integrating over $\varphi_1, \varphi_2$ and $x_2$ we finally have
\begin{equation}
S=\int dt\; \left[\frac{\ddot x_1+\gamma \dot x_1-f(x_1)}{g(x_1)}\right]^2,
\end{equation}
which is the usual action we would obtain ignoring the multiplicative character of the noise. We see that all the structure of Grassmann variables, which encodes the generalized stochastic $\alpha$ calculus, factors away; as we expected, since the solution of Eq.~\eqref{NonMarkSDE} can be directly performed and does not depend on any stochastic prescription. With this example we are showing that the preceding formalism, developed in principle to deal with Markov processes, is general enough to treat any system of SDE.

%%%%%%%%%%%%%%%%%%%%%%%
\subsection{Stochastic Landau-Lifshitz-Gilbert equation for micromagnetic dynamics}

Another very interesting application is a popular stochastic differential equation used  to describe dynamics of single magnetic moments in a ferromagnetic environment. The model  was proposed by Brown~\cite{Brown1963} and is  based on the classical Landau-Lifshitz-Gilbert equation (LLG)~\cite{Landau1935,Gilbert1955,Gilbert2004}.  

The adimensional  stochastic LLG  equation (sLLG) has the general form, 
\begin{equation}
\frac{d {\mathbf{m}}}{dt} =-
 \frac{1}{1+\eta^2} \ \mathbf{m} \times
 [(\mathbf{h}_{\eff}+ \mathbf{h}) + \eta \ {\mathbf{m}} \times(\mathbf{h}_{\eff}+ \mathbf{h} ) ]
 \; ,
 \label{eq:sLL1-adim}
\end{equation}
where ${\bf m}(t)$ is a single magnetization with constant modulus $|{\bf m}|^2=1$, $\mathbf{h}_{\eff}$ is some external magnetic field, $\eta$ is a dimensionless friction coefficient  and the Gaussian white noise ${\mathbf h}$ has zero average and correlations characterized by a 
diffusion constant $D$, 
\begin{equation}
\langle h_{i}(t) \rangle = 0
\;, \qquad
\langle  h_{i}(t)  h_{j}(t')  \rangle= 2 D \ \delta_{ij}  \delta(t-t')
\; .
\end{equation}
Clearly, Eq.~(\ref{eq:sLL1-adim}) is a system of SDE with  multiplicative noise, and  it should be interpreted in the sense of Stratonovich.  Of course, it is possible to rewrite the equation in any other interpretation by changing the drift using Eq.~(\ref{eq:Translator}) with $\alpha=1/2$. A complete $\alpha$-covariant formulation was recently presented in Ref.~\cite{Aron2014}. Moreover, numerical computations using different prescriptions were done in Ref.~\cite{Roma2014}.
 
Writing Eq.~(\ref{eq:sLL1-adim}) in the form of 
Eq.~(\ref{eq:LangevSystem}) we find, 
\begin{equation}
 \frac{dm_i(t)}{dt} = f_i({\bf m}) + g_{ij}({\bf m})\zeta_j(t) , 
\label{eq:LangevM}
\end{equation}
 where  $i,j = 1,2,3$ and  $\zeta_j(t)$ are $3$ independent Gaussian white noises,
\begin{equation}
\left\langle \zeta_i(t)\right\rangle   = 0 \;\;\mbox{,}\;\; 
\left\langle  \zeta_i(t), \zeta_j(t')\right\rangle = \delta_{ij} \delta(t-t').
\label{h-whitenoise}
\end{equation}
Notice that we have absorbed the diffusion constant $D$ into the definition of the matrix $\hat g$.
With this notation it is simple to identify the diffusion matrix and   
the drift. In fact, we find, 
\begin{equation}
g_{ij}=\frac{\sqrt{2D}}{1+\eta^2}\left(g^a_{ij}+g^s_{ij} \right)\; , 
\label{eq:g}
\end{equation}
with the symmetric and antisymmetric part given by 
\begin{eqnarray}
g_{ij}^a&=& \epsilon_{ijk} m_k ,
\label{eq:ga}\\
g_{ij}^s&=& \eta\left(m_im_j-m^2\delta_{ij}\right) \; .
\label{eq:gs}
\end{eqnarray}
The drift is given by 
\begin{equation}
f_i=\frac{1}{\sqrt{2D}} g_{ij} h_{{\rm \eff},j}.
\end{equation}

In this form, it is evident the transversal character of this equation since 
\begin{equation}
m_i g_{ij}=m_i f_i=0,
\end{equation}
and then,   
\begin{equation}
m_i\frac{d m_i}{dt}=0.
\end{equation}
This constraint implies that $|{\bf m}|^2=\mbox{constant}$ only in the Stratonovich interpretation, in which the chain rule is the usual one. 

With the Langevin structure given by Eq.~(\ref{eq:LangevM}), we immediately recognize that the generating functional should have the form of Eq.~(\ref{action}) or, upon integration over the Grassmann variables, it should have the form of Eq.~(\ref{eq:action-alpha}) with $\alpha=1/2$. 
By using Eqs.~(\ref{eq:g}),~(\ref{eq:ga}) and~(\ref{eq:gs}) to compute the several matrix products we find, 
 \begin{widetext}
\begin{eqnarray}
S & = & \int dt \left\{ 
\frac{D}{1+\eta^2}\left[\varphi^2 {\bf m}^2- ({\bf m}\cdot \varphi)^2 \right] + i\varphi_l(t) \left[ \partial_t m_l - \frac{1}{\sqrt{2D}} g_{lj} h_{{\rm \eff},j}  \right] \right.  \nonumber\\
& + & \left.  \frac{1}{2(1+\eta^2)}\left[ 
{\bf m}\cdot ({\bf \nabla}\times {\bf h}_{\rm eff}) +2\eta ({\bf m}\cdot {\bf h}_{\rm eff}) 
+\eta \left(m_im_j-m^2\delta_{ij}\right) \partial_i h_{{\rm eff}, j}\right] +  \frac{1}{4} \frac{D}{1+\eta^2} \mathbf{m}^2    \right\} \ .
\label{eq:action-m}
\end{eqnarray}
\end{widetext}
This result exactly coincides with the action presented in Ref.~\cite{Aron2014} in the case of $\alpha=1/2$, using a completely different construction.   

Notice that the quadratic term in the $\varphi$ variable is transversal. This fact imposes the constraint that $|{\bf m}(t)|=\mbox{constant}$.  
To gain more insight in the structure of this expression, it is instructive to look to the two-dimensional case. Physically, this is the case of two-dimensional rotor, submitted  to a three dimensional noisy magnetic field. This can be achieved by imposing to Eq.~(\ref{eq:action-m}) the constraint $m_3=0$. 

By making the following change of variables (with trivial Jacobian), 
\begin{eqnarray}
\left(
\begin{array}{c}
\varphi_1 \\ \varphi_2
\end{array}
\right)
=
\left(
\begin{array}{cc}
m_y & m_x \\ -m_x & m_y
\end{array}
\right)
\left(
\begin{array}{c}
\tilde\varphi_1 \\ \tilde \varphi_2
\end{array}
\right)\; , 
\end{eqnarray}
we can split the transversal and the longitudinal component of $\varphi$, in such a way that,  
\begin{eqnarray}
S & = & \int dt \frac{D}{1+\eta^2}\tilde\varphi^2_1 -i \tilde\varphi_1
 {\bf m}\times \left[\dot {\bf m} - {\bf f} ({\bf m}) \right]  \nonumber\\
& + & \tilde\varphi_2{\bf m}\cdot \dot{\bf m}+ \frac{1}{2}\left.\partial_k f_k({\bf m}) \right\} .
\end{eqnarray}
Note that there is no term proportional to $\tilde\varphi_2^2$. Then, integration over $\tilde\varphi_2$ simply imposes the constraint ${\bf m}\cdot d{\bf m}/dt=0$,  which can be solved using polar coordinates ${\bf m}= (\cos\theta, \sin\theta)$.
The generating functional in polar coordinates is given by 
\begin{equation}
Z=\int [D\theta] e^{-S[\theta]} ,
\end{equation}
with 
\begin{equation}
S =  \int dt \left[\frac{\dot {\theta} - f(\theta)}{\sigma} \right]^2+\frac{1}{2}\partial_\theta f(\theta) ,
\label{eq:actiontheta}
\end{equation}
where $\sigma=\sqrt{2D/(1+\eta^2)}$.

We can see that Eq.~(\ref{eq:actiontheta}) represents an additive process with effective diffusion 
constant $\sigma$.  Then, although in the original equation we have a multiplicative noise, the combined effect of transversality and dimensionality makes this process  essentially additive.  In three dimensions the result is very different. Although the dynamics is also transversal, the solution of the constraint in spherical coordinates is more involved and  it can be shown~\cite{Aron2014} that the stochastic process is truly multiplicative in three dimensions. 

The important point is that when dealing with the Grassmann functional formalism, all these results are automatically implemented by the Grassmann algebra and the structure of the diffusion matrix $\hat g$.

%%%%%%%%%%%%%%%%%%%%%%%%%
\section{Summary and discussion}
\label{sec:discusion}
%%%%%%%%%%%%%%%%%%%%%%%%%

We have discussed general multi-dimensional stochastic processes driven by a system of Langevin equations with multiplicative noise. We have addressed the problem arising from the variety of conventions available to deal with stochastic integrals and how the time reversal diffusion processes are affected by such conventions. In particular, we have developed some properties of the stochastic calculus in the $\alpha$ prescription by generalizing It\^o calculus. In appendix~\ref{sec:alphacalculus}, we carefully deduce some of these properties, specially the generalized chain rule, essential for any integration by parts. 

We have also presented a functional formalism to built up the generating functional of correlation functions in this type of processes. The generating functional is characterized by a functional integration over two sets of commuting variables as well as Grassmann variables. The advantage of this procedure is that we did not discretize the Langevin equations at any intermediate step, thus, we did not need to specify a particular stochastic prescription to built up the functional. As a consequence, the result of Eqs.~(\ref{eq:funcionalgenerador}) and~(\ref{action}) are completely general, representing a family of stochastic processes. In this representation, time reversal transformation became a linear transformation in the extended variables, simplifying in this way the complexity introduced by the mixture of prescriptions and the associated calculus rules. 
In some sense, the difficulties introduced by the generalized It\^o calculus are codified in our formalism in the Grassmann algebra. This fact allows us to compute any correlation function in a unified way.  
Of course, the problem of the specific stochastic prescription reappears if we decide to integrate out the Grassmann variables.  We performed this integration in section~\ref{sec:GrassmanIntegration} and it allowed us to identify the stochastic prescription with the definition of Grassmann Green's functions at the origin. It is very instructive to understand how the time reversed transformation $\alpha\to 1-\alpha$ appears in this formalism as a relation between the retarded and the advanced Green's function. These results completely generalize to a vector white-noise process the procedure presented in Refs.~\cite{arenas2010,arenas2012,Arenas2012-2} for a single variable.  

While this formalism shows its power to deal with Markov processes, where the problem of the stochastic prescriptions is essential, it is completely general and it is possible to use it for any stochastic process. To illustrate this point we presented some applications in section~\ref{sec:Examples}. We analyzed the explicit structure of the functional integration in the case of a non-Markov process driven by a second-order Langevin equation. In this case, the structure of the diffusion matrix together with the properties of Grassmann algebra completely factor out any dependence on the $\alpha$ parameter. Of  course, we recovered the known result that a stochastic non-Markov process in uniquely determined no matter if we deal with an additive or a multiplicative process.  We have also studied the functional representation of the stochastic LLG equation, used to describe micro magnetic dynamics. 
We built up a generating functional in the extended functional space and showed that, after integration over the Grassmann variables, the action coincides with a recently developed path integral formalism in the $\alpha$ prescription~\cite{Aron2014}. We have also focused in the two-dimensional case, and showed that the dynamical evolution of a two-dimensional rotor submitted to a three-dimensional noisy magnetic filed is in fact an {\em additive} stochastic process, very different from the three-dimensional case  which  is a truly multiplicative process. 
Again, all this information resides in the mathematical structure of the Grassmann variables.

Thus, having in hands this powerful formalism, we can face different applications. One of them is the formulation of fluctuation theorems satisfied by out-of-equilibrium multiplicative processes. Other interesting application, where this method is actually useful, is the study of out-of-equilibrium phase transitions driven by multiplicative noise. We are actually working on these subjects and we hope to report results soon.

\acknowledgments
The Brazilian agencies \emph{Conselho Nacional de Desenvolvimento 
Cient\'{\i}fico e Tecnol\'{o}gico (CNPq)} , \emph{Funda{\c{c}}{\~{a}}o de Amparo {\`{a}} 
Pesquisa do Estado do Rio de Janeiro (FAPERJ)} and \emph{ Coordena\c c\~ao de 
Aperfei\c coamento de Pessoal de N\'\i vel Superior (CAPES)} are acknowledged for
partial financial support. MVM is financed by a doctoral fellowship by
CAPES. Z.G.A. is financed by a post-doctoral fellowship by CAPES. 
D.G.B. is Senior Associate of the International Centre for Theoretical
Physics, ICTP, Trieste.

\appendix

%%%%%%%%%%%%%%%%
\section{Stochastic calculus in the $\alpha$-prescription}
\label{sec:alphacalculus}

%%%%%%%%%%%%%
\subsection{It\^o formula}
\label{sec:ItoFormula}

Most of the stochastic calculus is based on the notion that $dW$ is a differential of order $1/2$, which means that 
in taking limits retaining the order $dt$, expressions containing $dW^2$ should be considered. 
Usually this fact is summarized taking naively $dW^2=dt$. This intuition comes from the fact that a discretized Wiener process $W_i$, with $i=0,1,\ldots$,  satisfies  $ \langle \Delta W_i^2\rangle=\Delta t_i,$ where $\Delta W_i=
W_i-W_{i-1}$. However, the precise  meaning of $dW^2=dt$  is
\begin{equation}
\int_0^t [dW(t')]^2 G(t')=\int_0^t dt'  G(t') .
\label{dw2}
\end{equation}
This expression can be  rigorously  proved  provided we define the integrals in the  It\^o discretization and consider $G(t)$  in a class of functions usually called ``non-anticipating'' functions~\cite{gardiner}.  It is important to clarify these concepts to understand the demonstration and the range of applicability of the usual assumption $dW^2=dt$.
 
 To define the stochastic integrals we need to previously define the concept of mean square limit. We say that a sequence of stochastic variables $S_n$ converges to  $S$ ($\lim_{n\to\infty}S_n=S$) in the mean square sense, if 
\begin{equation}
\lim_{n\to \infty} \langle \left(S_n-S\right)^2\rangle=0 .
\end{equation}
With this definition the integral in the  It\^o prescription is
\begin{equation}
\int [dW(t')]^2 G(t')\equiv \lim_{n\to \infty} \sum_i G_{t_{i-1}} \Delta W_i^2 \ .
\label{dw2dt}
\end{equation}
To prove statement~\eqref{dw2}, let us define the integral 
\begin{equation}
I=\lim_{n\to\infty}\left\langle\left[ \sum_i G_{i-1} \Delta W_i^2-\sum_i G_{i-1}\Delta t_i\right]^2  \right\rangle.
\label{I}
\end{equation}
Clearly, if $I=0$, then the property (\ref{dw2}) is satisfied.  To proof that, the concept of non-anticipating function is central.  We say that  $G(t)$  is non-anticipating if  $G(t)$ and  $W(t')-W(t)$ are statistically uncorrelated for $t'> t$. So,
\begin{equation}
\langle G^p(t) (W(t')-W(t))^q\rangle=\langle G^p(t)\rangle\langle (W(t')-W(t))^q \rangle 
\label{nonanticipating}
\end{equation}
for any $p,q$ integers and $t'> t$. Using this property, we can write Eq.~(\ref{I}) as
\begin{eqnarray}
I&=&\lim_{n\to\infty}\left\{
\sum_i \langle G^2_{i-1}\rangle\langle (\Delta W_i^2-\Delta t_i)^2\rangle \right.
\nonumber \\
&+& 2 \left.\sum_{i<j} \langle G_{i-1}\rangle\langle (\Delta W_i^2-\Delta t_i)\rangle\langle  G_{j-1}\rangle\langle(\Delta W_j^2-\Delta t_j)\rangle\right\}. \nonumber \\
& &  \label{Iexpanded}
\end{eqnarray}
Using the usual properties of a Wiener process 
\begin{eqnarray}
& & \langle \Delta W_i^2\rangle=\Delta t_i,  \\
& & \langle (\Delta W_i^2-\Delta t_i)^2\rangle=2 \Delta t_i^2,
\end{eqnarray}
we immediately find that the last term of Eq.~(\ref{Iexpanded}) is zero and 
\begin{equation}
I=2 \lim_{n\to\infty} \left[
\sum_i  \Delta t_i^2 \langle G^2_{i-1}\rangle \right].
\end{equation}
Evidently, $I\to 0$ as $\Delta t_i ^2\to 0$. Then, 
\begin{equation}
\lim_{n\to\infty}\left[ \sum_i G_{i-1} \Delta W_i^2-\sum_i G_{i-1}\Delta t_i\right] =0,
\label{DeltaW2}
\end{equation}
and hence Eq.~(\ref{dw2dt}) is demonstrated. 

Thus, almost in any stochastic calculation,  we can safely replace $dW^2$ by $dt$, provided we are integrating non-anticipating functions with the It\^o interpretation.  Note that if the integral is defined in any other $\alpha$ prescription (even Stratonovich), the function $G(t)$ does not satisfy Eq.~(\ref{nonanticipating}) and consequently, Eq.~(\ref{DeltaW2}) cannot be proved. 
 
It is straightforward to extend this demonstration to the case of a n-dimensional Wiener process. In this case, if we deal with a set of $n$ Wiener processes $\{W_1(t),\ldots, W_n(t)\}$ we have
\begin{equation}
dW_i(t)dW_j(t)=\delta_{ij} dt,
\label{WWt}
\end{equation} 
where $i,j=1,\ldots, n$.

One immediate application of Eq.~(\ref{WWt}) is to compute a change of variables in a system of stochastic differential equations. 
Consider, for instance, a set of variables ${\bf x(t)}=\{x_1(t), \ldots, x_n(t)\}$ satisfying the differential stochastic equation in the It\^o prescription 
\begin{eqnarray}
 dx_i(t)= f_i({\bf x}(t)) dt + g_{ij}({\bf x}(t))dW_j(t), 
\label{dxito}
\end{eqnarray}
 where  $i = 1,\ldots,n$, $j = 1,\ldots,n$, ${\bf x} \in \Re^n$ and  $W_j(t)$ are $n$ independent Wiener processes.
The question is, what kind of differential equation obeys a set of function $F_s({\bf x(t)})$.
We can expand 
\begin{equation}
dF_s({\bf x}) = \partial_i F_s dx_i + \frac{1}{2}\partial_i\partial_j F_s dx_idx_j+\ldots  
\label{dFito}
\end{equation}
Introducing~(\ref{dxito}) into~(\ref{dFito}) and  retaining terms up to order $dt$, we find, 
\begin{eqnarray}
dF_s({\bf x})&=& \partial_i F_s \left[f_i({\bf x}(t)) dt + g_{ij}({\bf x}(t)\right]dW_j(t))  \nonumber \\
& &  +\frac{1}{2}\partial_i\partial_j F_s({\bf x}) g_{ik}({\bf x})g_{jl}({\bf x}) dW_kdW_l .
\end{eqnarray} 
Using Eq.~(\ref{WWt}), we immediately obtain the celebrated  It\^o formula
\begin{eqnarray}
dF_s({\bf x})&=& \left(\partial_i F_s({\bf x}) f_i({\bf x})+\frac{1}{2}\partial_i\partial_j F_s({\bf x}) g_{ik}({\bf x})g_{jk}({\bf x}) \right)   dt  \nonumber \\
& &  +\partial_i F_s({\bf x}) g_{ij}({\bf x}) dW_j(t)\ .
\label{ItoFormula}
\end{eqnarray}

%%%%%%%%%%%%%%%%%%%%%%%%%%
\subsection{Translations between arbitrary prescriptions $\alpha\to \beta$}
\label{Ap:alphabeta}
%%%%%%%%%%%%%%%%%%%%%%%%%
Using the It\^o formula,  deduced in the preceding section, we can build a translation rule, very useful to describe a particular stochastic process in any convenient prescription $\alpha$. 
For concreteness, 
consider  a stochastic process that is described by a  SDE in the $\alpha$ interpretation. Then,  it is always possible 
to find \emph{another SDE}  in another prescription, say $\beta$, that has exactly the same solution, \emph{i.e.}, represents the same stochastic process. 

Consider, for instance, a SDE given by 
\begin{equation}
dx_i=f_i dt +g_{ij} dW_j
\label{eqito}
\end{equation} 
in the It\^o interpretation. 
The question is, is it possible to write an SDE
\begin{equation}
dx_i=a_i dt +b_{ij} dW_j
\label{eqalpha}
\end{equation} 
in the $\alpha$ interpretation with the same solution?  To answer this question, we 
integrate Eq.~(\ref{eqalpha})
\begin{equation}
x_i(t)=\int  a_i dt + \int_\alpha  b_{ij} dW_j,
\end{equation} 
 where the last integral means 
 \begin{equation}
 \int_\alpha  b_{ij} dW_j=\lim_{n\to\infty} \sum_k b_{ij}[(1-\alpha) x_{k-1}+\alpha x_k]\Delta_k W_j.
 \label{intalpha}
 \end{equation}
 Expanding the function $b_{ij}$
 \begin{eqnarray}
 \lefteqn{b_{ij}[(1-\alpha) x_{k-1}+\alpha x_k]  =  b_{ij}[x_{k-1}+\alpha dx_k]} \nonumber \\
& = & b_{ij}[x_{k-1}] +\alpha \partial_sb_{ij}[x_{k-1}] dx_s.
\label{bexp}
 \end{eqnarray}
 Introducing in Eq.~(\ref{bexp}) the expression for $dx_s$ given by Eq.~(\ref{eqito})  
\begin{eqnarray}
 \lefteqn{b_{ij}[(1-\alpha) x_{k-1}+\alpha x_k]  =  b_{ij}[x_{k-1}]} \nonumber \\
&& \ + \alpha\partial_s b_{ij}[x_{k-1}] (f_s \Delta t +g_{sk} \Delta W_k) 
\label{00}
 \end{eqnarray}
and  retaining terms proportional to $\Delta W_i\Delta W_j=\delta_{ij} dt$ when computing  Eq.~(\ref{intalpha})
we find 
 \begin{equation}
 \int_\alpha  b_{ij} dW_j=\int_0  b_{ij} dW_j + \alpha \int  \partial_s b_{ij} g_{sj} dt,
 \label{intalphazero}
 \end{equation}
where $\int_\alpha$ and $\int_0$ mean integration in the $\alpha$ and It\^o prescription, respectively. Here it is interesting to observe that the connection between  integrals in the $\alpha$ and It\^o prescription is due to the fact that the stochastic variable $x(t)$ satisfies an It\^o SDE. In general, integrals of arbitrary functions in different prescriptions are completely uncorrelated.

Introducing Eq.~(\ref{intalphazero})  into Eq.~(\ref{eqalpha}),
\begin{equation}
x=\int  (a_i+\alpha \partial_s b_{ij} g_{sj} )dt+ \int_0 b_{ij} dW_j .
\end{equation}
Comparing with Eq.~(\ref{eqito}) we see that 
\begin{eqnarray}
f_i&=&a_i+\alpha g_{kj}\partial_k g_{ij},   \\
g_{ij}&=&b_{ij}.
\end{eqnarray}
Then, we have the following relations for the SDE in the $\alpha$ and Ito prescriptions,
\[
\begin{array}{|c|c|}
\hline
\mbox{It\^o SDE} &  dx_i=f_i dt + g_{ij} dW_j  \\
\hline
\alpha \mbox{SDE} &  dx_i=(f_i-\alpha g_{kj}\partial_k g_{ij})+ g_{ij} dW_j   \\
\hline
\end{array}
\]
or, equivalently,
\[
\begin{array}{|c|c|}
\hline
\alpha \mbox{SDE} &  dx_i=f_i dt + g_{ij} dW_j  \\
\hline
 \mbox{It\^o SDE} &  dx_i=(f_i+\alpha g_{kj}\partial_k g_{ij})+ g_{ij} dW_j   \\
\hline
\end{array}
\]

Thus, we can transform  from a SDE in the $\alpha$ prescription to another one in the It\^o interpretation and  vice versa. Then, we can use transitivity to go from  $\alpha$ to a general 
$\beta$ prescription obtaining, 
\[
\begin{array}{|c|c|}
\hline
\alpha \mbox{SDE} &  dx_i=f_i dt + g_{ij} dW_j  \\
\hline
\beta \mbox{SDE} &  dx_i=(f_i+(\alpha-\beta) g_{kj}\partial_k g_{ij})+ g_{ij} dW_j   \\
\hline
\end{array}
\]

%%%%%%%%%%%%%%%%%%%%%%%%%%%%
\subsection{Chain Rule: single variable}
\label{Ap:chainrule1}
%%%%%%%%%%%%%%%%%%%%%%%%%%%%
Integration by parts and the derivatives of composed functions, the ``chain rule'', are very useful and important related concepts. Here, we show the form of the chain rule in the stochastic $\alpha$ prescription calculus. 

Consider the following SDE interpreted in the $\alpha$-prescription
\begin{equation}
 dx=f(x) dt + g(x) dW.
\end{equation}
Using the results of the preceding section, it is simple to write a SDE interpreted in the It\^o prescription and with the 
same solution, that is,
\begin{equation}
 dx=(f(x)+\alpha g g' ) dt + g(x) dW.
\end{equation}

We now perform the change of variables $y=F(x)$ and use the It\^o formula to write
\begin{eqnarray}
 dy&=&\left[ F'(x) f(x)+\alpha g g' F'(x)+\frac{1}{2} g^2 F''(x)\right] dt 
\nonumber \\ 
& & \ + g(x)F'(x) dW.
\end{eqnarray}
We can rewrite this equation in terms of the new variable $y$, defining the 
inverse function $G(y)$ in the following form 
\begin{eqnarray}
&& y=F(x),  ~~~~~~~ x= G(y),  \\
&& F'(x)= (G')^{-1}(y),  \\
&& F''(x)=- G''(y) (G')^{-3}(y).
\end{eqnarray}
With this,
\begin{eqnarray}
 dy & = & \left[(G')^{-1} \tilde f  +\alpha (G')^{-2} \tilde g \tilde g' 
-\frac{1}{2} \tilde g^2 G''(G')^{-3}\right] dt \nonumber \\ 
& & \ +   (G')^{-1} \tilde g dW,
\end{eqnarray}
where we have defined $\tilde f(y)=f(G(y))$ and 
$\tilde g(y)=g(G(y))$.
Thus, we have a SDE in the It\^o prescription of the form 
\begin{equation}
dy=a(y) dt + b(y) dW.
\end{equation}
Now,  we can return to the alpha prescription making
\begin{equation}
dy=(a(y)-\alpha b b' dt) + b(y) dW,
\end{equation}
with $b= (G')^{-1} \tilde g$.
\begin{eqnarray}
dy & = & \left[\tilde f(y) dt +\tilde g(y) dW \right] (G')^{-1}(y) \nonumber \\
& & \ + \left(\alpha-\frac{1}{2}\right)\tilde g ^2 (G')^{-3} G'' dt.
\end{eqnarray}
Turning back to the original  $x$ variable  we find
\begin{eqnarray}
 dF(x)&=&\left[f(x) dt +g(x) dW \right] F'(x) \nonumber \\
& & \ + \left(\frac{1}{2}-\alpha\right) g^2(x) F''(x) .
\end{eqnarray}
From this equation we immediately deduce the generalized chain rule
\begin{equation}
\frac{dF}{dt}= F'(x)\frac{dx}{dt}+  \left(\frac{1-2 \alpha}{2}\right) g^2(x) 
F''(x).
\end{equation}
For the Stratonovich convention, $\alpha=1/2$, this rule is the usual one,  while for $\alpha=0$ this equation is a very well known consequence of the It\^o formula, Eq.~(\ref{ItoFormula}). 

%%%%%%%%%%%%%%%%%%%%%%%%%%%%%%%
\subsection{Chain rule: multiple variables}
\label{Ap:chainrule2}
%%%%%%%%%%%%%%%%%%%%%%%%%%%%%%%%%

The chain rule in the $\alpha$ prescription, demonstrated in the previous section, can be easily generalized to the case of a multivariate problem in which we have a set of $n$ independent Wiener processes. Although the matrix algebra is slightly more involved, the general steps of the demonstration are the same as in the one variable case. 

Consider the following system of SDE interpreted in the $\alpha$-prescription
\begin{equation}
 dx_i=f_i({\bf x}) dt + g_{ij}({\bf x}) dW_j.
\end{equation}
It is simple to write this system in the It\^o prescription having, of course,  the 
same solution. That is, 
\begin{equation}
 dx_i=(f_i({\bf x})+\alpha g_{k\ell}\partial_k g_{i\ell} ) dt + g({\bf x})_{ij} dW_j.
\end{equation}
Now we perform the change of variables $y_s=F_s({\bf x})$ and use the It\^o formula to write
\begin{eqnarray}
 dy_s&=&\left[ \partial_iF_s({\bf x})\left( f_i({\bf x})+\alpha g_{k\ell} \partial_k g_{i\ell}\right) 
 +\frac{1}{2} g_{ik}g_{jk} \partial_i\partial_jF({\bf x})\right] dt \nonumber \\
 & & \ + g_{ij}({\bf x})\partial_i F_s({\bf x}) dW_j.
\end{eqnarray}
We can rewrite this equation in terms of the new variables $y_s$, defining the 
inverse set of functions $G_s(y)$ in the following form 
\begin{eqnarray}
&& y_s=F_s({\bf x}),  ~~~~~~~ x_s= G_s({\bf y}),  \\
&& \partial_i F_s({\bf x})=\left(\partial_i G_s\right)^{-1}(y)\equiv (\partial G)_{is}^{-1}(y),  \\
&& \partial_i\partial_jF_s({\bf x})=- \left(\partial G\right)^{-2}_{sk}\left(\partial G\right)^{-1}_{\ell j} \partial_\ell \partial_k G_i.
\end{eqnarray}

With this,
\begin{equation}
dy_s=a_s({\bf y}) dt + b_{ij}({\bf y}) dW_j,
\end{equation}
with 
\begin{eqnarray}
a_s({\bf y})&=&(\partial G)_{is}^{-1} \left(\tilde f_i(y) -\alpha (\partial G)^{-1}_{pk} \tilde g_{k\ell}  \partial_p 
\tilde g_{i\ell}\right) 
\nonumber \\
 & & \ - \frac{1}{2} \tilde g_{ik}  \tilde g_{jk} \left(\partial G\right)_{sp}^{-2}\left(\partial G\right)^{-1}_{lj} \partial_\ell\partial_p G_i , \\
b_{ij}({\bf y})&=& \left(\partial G\right)_{is}^{-1}({\bf y})  \tilde g_{ij}({\bf y}), 
\end{eqnarray}
where we have defined $\tilde f_i({\bf y})=f_i({\bf G}({\bf y}))$ and 
$\tilde g_{ij}({\bf y})=g_{ij}({\bf G}({\bf y}))$.
 Now,  we can return to the $\alpha$ prescription by making
\begin{equation}
dy_s=\left(a_s({\bf y})-\alpha b_{k\ell} \partial_k b_{s\ell} \right)dt + b_{s\ell}({\bf y}) dW_{\ell},
\end{equation}
obtaining 
\begin{eqnarray}
dy_s&=&\left[\tilde f_i({\bf y}) dt +\tilde g_{ij}({\bf y}) dW_j \right] (\partial G)_{is}^{-1}({\bf y})  \\
&+& \left(\frac{1}{2}-\alpha\right) \tilde g_{ik}  \tilde g_{jk} \left(\partial G\right)_{sp}^{-2}\left(\partial G\right)^{-1}_{lj} \partial_\ell\partial_p G_i  dt. \nonumber 
\end{eqnarray}
Turning back to the original  $x$ variable we find
\begin{eqnarray}
 dF_s({\bf x})&=&\left[f_i({\bf x}) dt +g_{ij}({\bf x}) dW_j \right] \partial_i F_s({\bf x}) \nonumber \\
&& \ + \left(\frac{1}{2}-\alpha\right) g_{ik}({\bf x})g_{jk}({\bf x}) \partial_i\partial_j F_s({\bf x}) 
\end{eqnarray}
From this equation, we immediately deduce the generalized chain rule,
\begin{equation}
\frac{dF({\bf x})}{dt}= \partial_i F({\bf x})\frac{dx_i}{dt}+  \left(\frac{1-2 \alpha}{2}\right) g_{ik}({\bf x})g_{jk}({\bf x}) \partial_i\partial_j F({\bf x}).
\end{equation}

%%%%%%%%%%%%%%%%%%%%%%%%%%%%
\section{Determinants and  Grassmann variables}
\label{sec:Determinants}
%%%%%%%%%%%%%%%%%%%%%%%%%%%%%

In order to make the paper self-contained, in this appendix we briefly review some usual manipulations with Grassmann variables~\cite{arenas2010}.  
For a detailed treatment of this subject in statistical mechanics as well as in quantum field theory we refer the reader to Ref.~\onlinecite{Zinn-Justin}. 

We define a set of $n$ Grassmann variables $\theta_i$ and its conjugates $\bar\theta_i$, where $i=1\ldots n$ as 
\begin{equation}
\{\theta_i,\theta_j\}=\{\bar\theta_i,\bar\theta_j\}=\{\theta_i,\bar\theta_j\}=0 
\end{equation}  
where $\{a,b\}=ab+ba$ is the anti-commutator of $a$ and $b$. This definition implies the nilpotent property $\theta_i^2=\bar\theta_i^2=0$. Therefore, any function of these variables should be a polynomial of degree at least $2n$. For instance, if $n=1$, 
\begin{equation}
F(\theta,\bar\theta)=A + B\;\theta + C\;\bar\theta + D\;\theta\bar\theta
\label{Fn1}
\end{equation}  
is the most general function of $\theta$ and $\bar\theta$. The coefficients $A,B,C,D$ are  complex numbers. 

Differentiation in the Grassmann variable is also a nilpotent operator $\partial^2/\partial^2\theta=0$ satisfying the Clifford algebra
\begin{equation}
\{\frac{\partial~}{\partial\theta_i},\frac{\partial~}{\partial\theta_j}\}=0,\;\;\; 
\{\frac{\partial~}{\partial\theta_i},\theta_j\}=\delta_{ij} 
\end{equation}  
and its conjugates. Any anti-commutator that mixes $\theta_i$ and $\bar\theta_j$ is also zero since they are independent. 

Interestingly, integration in a Grassmann space is the same operation as differentiation. Therefore, the use of a derivative or an integral symbol is a matter of taste. For instance, taking into account Eq.~(\ref{Fn1}),  
\begin{equation}
\int d\theta d\bar\theta\; F(\theta, \bar\theta)=D
\label{integral}
\end{equation}
That means that the integration over $d\theta d\bar\theta$ picks up the coefficient of the term $\theta\bar\theta$. 

Let us consider the Gaussian Grassmann integral 
\begin{equation}
I_G(A)=\int \left(\prod_{k=1}^{n}d\theta_k d\bar\theta_k\right)\;\;
e^{\sum_{ij}\bar\theta_i A_{ij} \theta_j }
\label{IGA}
\end{equation}
where $A_{ij}$ are the elements of a matrix $A$.
According with Eq.~(\ref{integral}), the integral is the coefficient of the term proportional to $\theta_1\bar\theta_1\theta_2\bar\theta_2\ldots\theta_n\bar\theta_n$. 
Therefore, expanding the exponential in Eq.~(\ref{IGA}) in a ``finite'' Taylor series, and reordering the terms taking into account the anti-commuting properties of the Grassmann variables, 
\begin{equation}
I_G(A)=\sum_{j_1,j_2,\ldots, j_n} (-1)^{P} A_{n,j_n}A_{n-1,j_{n-1}}\ldots A_{1,j_1}
\end{equation}
where $P$ is the order of the permutation of $\{j_1,\ldots, j_n\}$. We immediately recognize that 
\begin{equation}
I_G(A)=\det(A)
\label{detA}
\end{equation}
This result should be compare with the output of a normal Gaussian integral that is $I(A)\sim {\det^{-1}(A)}$. Therefore, 
in the same way that the inverse of an  $n\times n$ matrix determinant  can be represented as a Gaussian integral over a set of $n$ complex variables, the determinant itself can be represented as a Gaussian integral over a  set of $n$ complex Grassmann variables.

We can generalize now the case of a discrete set of Grassmann variables $\{\theta_1,\ldots\theta_n\}$, to an infinite set of continuous variables; {\it i.\ e.\ }, a Grassmann function $\xi(t)$.  In this case, we can generalize Eqs.~(\ref{IGA}) and~(\ref{detA}) to
\begin{equation}
\det(A)=\int \mathit{D}\xi \mathit{D}\bar{\xi} \ e^{\int dt dt'\;\bar{\xi}(t) A(t,t')\xi(t')}
\end{equation} 
where $\det(A)$ is a functional determinant and $\mathit{D}\xi \mathit{D}\bar{\xi}$ are functional integrations over Grassmann variables.
$A(t,t')$ is the kernel of the functional $A$. 
This is the formula used in Eq.~(\ref{detgrassman}) to represent the Jacobian $\delta\hat O/\delta x$.

%\bibliography{stochastic}

\begin{thebibliography}{39}%
\makeatletter
\providecommand \@ifxundefined [1]{%
 \@ifx{#1\undefined}
}%
\providecommand \@ifnum [1]{%
 \ifnum #1\expandafter \@firstoftwo
 \else \expandafter \@secondoftwo
 \fi
}%
\providecommand \@ifx [1]{%
 \ifx #1\expandafter \@firstoftwo
 \else \expandafter \@secondoftwo
 \fi
}%
\providecommand \natexlab [1]{#1}%
\providecommand \enquote  [1]{``#1''}%
\providecommand \bibnamefont  [1]{#1}%
\providecommand \bibfnamefont [1]{#1}%
\providecommand \citenamefont [1]{#1}%
\providecommand \href@noop [0]{\@secondoftwo}%
\providecommand \href [0]{\begingroup \@sanitize@url \@href}%
\providecommand \@href[1]{\@@startlink{#1}\@@href}%
\providecommand \@@href[1]{\endgroup#1\@@endlink}%
\providecommand \@sanitize@url [0]{\catcode `\\12\catcode `\$12\catcode
  `\&12\catcode `\#12\catcode `\^12\catcode `\_12\catcode `\%12\relax}%
\providecommand \@@startlink[1]{}%
\providecommand \@@endlink[0]{}%
\providecommand \url  [0]{\begingroup\@sanitize@url \@url }%
\providecommand \@url [1]{\endgroup\@href {#1}{\urlprefix }}%
\providecommand \urlprefix  [0]{URL }%
\providecommand \Eprint [0]{\href }%
\providecommand \doibase [0]{http://dx.doi.org/}%
\providecommand \selectlanguage [0]{\@gobble}%
\providecommand \bibinfo  [0]{\@secondoftwo}%
\providecommand \bibfield  [0]{\@secondoftwo}%
\providecommand \translation [1]{[#1]}%
\providecommand \BibitemOpen [0]{}%
\providecommand \bibitemStop [0]{}%
\providecommand \bibitemNoStop [0]{.\EOS\space}%
\providecommand \EOS [0]{\spacefactor3000\relax}%
\providecommand \BibitemShut  [1]{\csname bibitem#1\endcsname}%
\let\auto@bib@innerbib\@empty
%</preamble>
\bibitem [{\citenamefont {Gardiner}(1996)}]{gardiner}%
  \BibitemOpen
  \bibfield  {author} {\bibinfo {author} {\bibfnamefont {C.~W.}\ \bibnamefont
  {Gardiner}},\ }\href@noop {} {\emph {\bibinfo {title} {Handbook of stochastic
  methods for physics, chemistry and the natural sciences}}}\ (\bibinfo
  {publisher} {Springer-Verlag},\ \bibinfo {address} {Berlin Heidelberg},\
  \bibinfo {year} {1996})\BibitemShut {NoStop}%
\bibitem [{\citenamefont {van Kampen}(2007)}]{vanKampen}%
  \BibitemOpen
  \bibfield  {author} {\bibinfo {author} {\bibfnamefont {N.~G.}\ \bibnamefont
  {van Kampen}},\ }\href@noop {} {\emph {\bibinfo {title} {Stochastic Processes
  in Physics and Chemistry}}}\ (\bibinfo  {publisher} {Elsevier},\ \bibinfo
  {address} {London, UK},\ \bibinfo {year} {2007})\BibitemShut {NoStop}%
\bibitem [{\citenamefont {Crooks}(1999)}]{crooks1999}%
  \BibitemOpen
  \bibfield  {author} {\bibinfo {author} {\bibfnamefont {G.~E.}\ \bibnamefont
  {Crooks}},\ }\href {\doibase 10.1103/PhysRevE.60.2721} {\bibfield  {journal}
  {\bibinfo  {journal} {Phys. Rev. E}\ }\textbf {\bibinfo {volume} {60}},\
  \bibinfo {pages} {2721} (\bibinfo {year} {1999})}\BibitemShut {NoStop}%
\bibitem [{\citenamefont {Seifert}(2005)}]{seifert2005}%
  \BibitemOpen
  \bibfield  {author} {\bibinfo {author} {\bibfnamefont {U.}~\bibnamefont
  {Seifert}},\ }\href {\doibase 10.1103/PhysRevLett.95.040602} {\bibfield
  {journal} {\bibinfo  {journal} {Phys. Rev. Lett.}\ }\textbf {\bibinfo
  {volume} {95}},\ \bibinfo {pages} {040602} (\bibinfo {year}
  {2005})}\BibitemShut {NoStop}%
\bibitem [{\citenamefont {Seifert}(2008)}]{seifert2008}%
  \BibitemOpen
  \bibfield  {author} {\bibinfo {author} {\bibfnamefont {U.}~\bibnamefont
  {Seifert}},\ }\href {http://dx.doi.org/10.1140/epjb/e2008-00001-9} {\bibfield
   {journal} {\bibinfo  {journal} {EPJ B}\ }\textbf {\bibinfo {volume} {64}},\
  \bibinfo {pages} {423} (\bibinfo {year} {2008})}\BibitemShut {NoStop}%
\bibitem [{\citenamefont {Martin}\ \emph {et~al.}(1973)\citenamefont {Martin},
  \citenamefont {Siggia},\ and\ \citenamefont {Rose}}]{MSR1973}%
  \BibitemOpen
  \bibfield  {author} {\bibinfo {author} {\bibfnamefont {P.~C.}\ \bibnamefont
  {Martin}}, \bibinfo {author} {\bibfnamefont {E.~D.}\ \bibnamefont {Siggia}},
  \ and\ \bibinfo {author} {\bibfnamefont {H.~A.}\ \bibnamefont {Rose}},\
  }\href {\doibase 10.1103/PhysRevA.8.423} {\bibfield  {journal} {\bibinfo
  {journal} {Phys. Rev. A}\ }\textbf {\bibinfo {volume} {8}},\ \bibinfo {pages}
  {423} (\bibinfo {year} {1973})}\BibitemShut {NoStop}%
\bibitem [{\citenamefont {Janssen}(1976)}]{Janssen1976}%
  \BibitemOpen
  \bibfield  {author} {\bibinfo {author} {\bibfnamefont {H.-K.}\ \bibnamefont
  {Janssen}},\ }\href {\doibase 10.1007/BF01316547} {\bibfield  {journal}
  {\bibinfo  {journal} {Z. Phys B}\ }\textbf {\bibinfo {volume} {23}},\
  \bibinfo {pages} {377} (\bibinfo {year} {1976})}\BibitemShut {NoStop}%
\bibitem [{\citenamefont {{C. De Dominics}}(1976)}]{deDominicis}%
  \BibitemOpen
  \bibfield  {author} {\bibinfo {author} {\bibnamefont {{C. De Dominics}}},\
  }\href {\doibase 10.1051/jphyscol:1976138} {\bibfield  {journal} {\bibinfo
  {journal} {J. Phys. Colloques}\ }\textbf {\bibinfo {volume} {37}},\ \bibinfo
  {pages} {C1 247} (\bibinfo {year} {1976})}\BibitemShut {NoStop}%
\bibitem [{\citenamefont {Freund}\ and\ \citenamefont
  {P{\"o}schel}(2000)}]{Poschel}%
  \BibitemOpen
  \bibfield  {author} {\bibinfo {author} {\bibfnamefont {J.~A.}\ \bibnamefont
  {Freund}}\ and\ \bibinfo {author} {\bibfnamefont {T.}~\bibnamefont
  {P{\"o}schel}},\ }\href@noop {} {\emph {\bibinfo {title} {Stochastic
  Processes in Physics, Chemistry and Biology}}}\ (\bibinfo  {publisher}
  {Springer-Verlag},\ \bibinfo {address} {Berlin, Heidelberg},\ \bibinfo {year}
  {2000})\BibitemShut {NoStop}%
\bibitem [{\citenamefont {Murray}(2002)}]{Murray}%
  \BibitemOpen
  \bibfield  {author} {\bibinfo {author} {\bibfnamefont {J.~D.}\ \bibnamefont
  {Murray}},\ }\href@noop {} {\emph {\bibinfo {title} {Mathematical Biology. I.
  An introduction}}}\ (\bibinfo  {publisher} {Springer-Verlag},\ \bibinfo
  {address} {Berlin, Heidelberg},\ \bibinfo {year} {2002})\BibitemShut
  {NoStop}%
\bibitem [{\citenamefont {Mantegna}\ and\ \citenamefont
  {Stanley}(2000)}]{Mantegna}%
  \BibitemOpen
  \bibfield  {author} {\bibinfo {author} {\bibfnamefont {R.~N.}\ \bibnamefont
  {Mantegna}}\ and\ \bibinfo {author} {\bibfnamefont {H.~E.}\ \bibnamefont
  {Stanley}},\ }\href@noop {} {\emph {\bibinfo {title} {An introduction to
  econophysics: correlations and complexity in finance}}}\ (\bibinfo
  {publisher} {Cambridge University Press},\ \bibinfo {address} {Cambridge,
  UK},\ \bibinfo {year} {2000})\BibitemShut {NoStop}%
\bibitem [{\citenamefont {Bouchaud}\ and\ \citenamefont
  {Potters}(2003)}]{Bouchaud}%
  \BibitemOpen
  \bibfield  {author} {\bibinfo {author} {\bibfnamefont {J.~P.}\ \bibnamefont
  {Bouchaud}}\ and\ \bibinfo {author} {\bibfnamefont {M.}~\bibnamefont
  {Potters}},\ }\href@noop {} {\emph {\bibinfo {title} {Theory of financial
  risk and derivative pricing: from statistical physics to risk management}}}\
  (\bibinfo  {publisher} {Cambridge University Press},\ \bibinfo {year}
  {2003})\BibitemShut {NoStop}%
\bibitem [{\citenamefont {Lan\c{c}on}\ \emph {et~al.}(2001)\citenamefont
  {Lan\c{c}on}, \citenamefont {Batrouni}, \citenamefont {Lobry},\ and\
  \citenamefont {Ostrowsky}}]{Lancon2001}%
  \BibitemOpen
  \bibfield  {author} {\bibinfo {author} {\bibfnamefont {P.}~\bibnamefont
  {Lan\c{c}on}}, \bibinfo {author} {\bibfnamefont {G.}~\bibnamefont
  {Batrouni}}, \bibinfo {author} {\bibfnamefont {L.}~\bibnamefont {Lobry}}, \
  and\ \bibinfo {author} {\bibfnamefont {N.}~\bibnamefont {Ostrowsky}},\ }\href
  {http://stacks.iop.org/0295-5075/54/i=1/a=028} {\bibfield  {journal}
  {\bibinfo  {journal} {Europhys. Lett.}\ }\textbf {\bibinfo {volume} {54}},\
  \bibinfo {pages} {28} (\bibinfo {year} {2001})}\BibitemShut {NoStop}%
\bibitem [{\citenamefont {Lan\c{c}on}\ \emph {et~al.}(2002)\citenamefont
  {Lan\c{c}on}, \citenamefont {Batrouni}, \citenamefont {Lobry},\ and\
  \citenamefont {Ostrowsky}}]{Lancon2002}%
  \BibitemOpen
  \bibfield  {author} {\bibinfo {author} {\bibfnamefont {P.}~\bibnamefont
  {Lan\c{c}on}}, \bibinfo {author} {\bibfnamefont {G.}~\bibnamefont
  {Batrouni}}, \bibinfo {author} {\bibfnamefont {L.}~\bibnamefont {Lobry}}, \
  and\ \bibinfo {author} {\bibfnamefont {N.}~\bibnamefont {Ostrowsky}},\ }\href
  {http://www.sciencedirect.com/science/article/pii/S0378437101005106}
  {\bibfield  {journal} {\bibinfo  {journal} {Physica A}\ }\textbf {\bibinfo
  {volume} {304}},\ \bibinfo {pages} {65 } (\bibinfo {year}
  {2002})}\BibitemShut {NoStop}%
\bibitem [{\citenamefont {Garc\'ia-Palacios}\ and\ \citenamefont
  {L\'azaro}(1998)}]{Palacios1998}%
  \BibitemOpen
  \bibfield  {author} {\bibinfo {author} {\bibfnamefont {J.~L.}\ \bibnamefont
  {Garc\'ia-Palacios}}\ and\ \bibinfo {author} {\bibfnamefont {F.~J.}\
  \bibnamefont {L\'azaro}},\ }\href {\doibase 10.1103/PhysRevB.58.14937}
  {\bibfield  {journal} {\bibinfo  {journal} {Phys. Rev. B}\ }\textbf {\bibinfo
  {volume} {58}},\ \bibinfo {pages} {14937} (\bibinfo {year}
  {1998})}\BibitemShut {NoStop}%
\bibitem [{\citenamefont {G.~Bertotti}\ and\ \citenamefont
  {Serpico}(2009)}]{Bertottl}%
  \BibitemOpen
  \bibfield  {author} {\bibinfo {author} {\bibfnamefont {I.~M.}\ \bibnamefont
  {G.~Bertotti}}\ and\ \bibinfo {author} {\bibfnamefont {C.}~\bibnamefont
  {Serpico}},\ }\href@noop {} {\emph {\bibinfo {title} {{Nonlinear
  Magnetization Dynamics in Nanosystems}}}}\ (\bibinfo  {publisher}
  {Elsevier},\ \bibinfo {address} {Amsterdam, The Netherlands},\ \bibinfo
  {year} {2009})\BibitemShut {NoStop}%
\bibitem [{\citenamefont {Janssen}(1992)}]{Janssen-RG}%
  \BibitemOpen
  \bibfield  {author} {\bibinfo {author} {\bibfnamefont {H.~K.}\ \bibnamefont
  {Janssen}},\ }\href@noop {} {\emph {\bibinfo {title} {From phase transitions
  to chaos: Topics in Modern Statistical Physics}}}\ (\bibinfo  {publisher}
  {World Scientific},\ \bibinfo {address} {Singapore},\ \bibinfo {year}
  {1992})\BibitemShut {NoStop}%
\bibitem [{\citenamefont {Aron}\ \emph {et~al.}(2010)\citenamefont {Aron},
  \citenamefont {Biroli},\ and\ \citenamefont {Cugliandolo}}]{AronLeticia2010}%
  \BibitemOpen
  \bibfield  {author} {\bibinfo {author} {\bibfnamefont {C.}~\bibnamefont
  {Aron}}, \bibinfo {author} {\bibfnamefont {G.}~\bibnamefont {Biroli}}, \ and\
  \bibinfo {author} {\bibfnamefont {L.~F.}\ \bibnamefont {Cugliandolo}},\
  }\href {http://stacks.iop.org/1742-5468/2010/i=11/a=P11018} {\bibfield
  {journal} {\bibinfo  {journal} {J. Stat. Mech.}\ ,\ \bibinfo {pages}
  {P11018}} (\bibinfo {year} {2010})}\BibitemShut {NoStop}%
\bibitem [{\citenamefont {Arenas}\ and\ \citenamefont
  {Barci}(2010)}]{arenas2010}%
  \BibitemOpen
  \bibfield  {author} {\bibinfo {author} {\bibfnamefont {Z.~G.}\ \bibnamefont
  {Arenas}}\ and\ \bibinfo {author} {\bibfnamefont {D.~G.}\ \bibnamefont
  {Barci}},\ }\href {\doibase 10.1103/PhysRevE.81.051113} {\bibfield  {journal}
  {\bibinfo  {journal} {Phys. Rev. E}\ }\textbf {\bibinfo {volume} {81}},\
  \bibinfo {pages} {051113} (\bibinfo {year} {2010})}\BibitemShut {NoStop}%
\bibitem [{\citenamefont {Arenas}\ and\ \citenamefont
  {Barci}(2012{\natexlab{a}})}]{arenas2012}%
  \BibitemOpen
  \bibfield  {author} {\bibinfo {author} {\bibfnamefont {Z.~G.}\ \bibnamefont
  {Arenas}}\ and\ \bibinfo {author} {\bibfnamefont {D.~G.}\ \bibnamefont
  {Barci}},\ }\href {\doibase 10.1103/PhysRevE.85.041122} {\bibfield  {journal}
  {\bibinfo  {journal} {Phys. Rev. E}\ }\textbf {\bibinfo {volume} {85}},\
  \bibinfo {pages} {041122} (\bibinfo {year} {2012}{\natexlab{a}})}\BibitemShut
  {NoStop}%
\bibitem [{\citenamefont {Arenas}\ and\ \citenamefont
  {Barci}(2012{\natexlab{b}})}]{Arenas2012-2}%
  \BibitemOpen
  \bibfield  {author} {\bibinfo {author} {\bibfnamefont {Z.~G.}\ \bibnamefont
  {Arenas}}\ and\ \bibinfo {author} {\bibfnamefont {D.~G.}\ \bibnamefont
  {Barci}},\ }\href {http://stacks.iop.org/1742-5468/2012/i=12/a=P12005}
  {\bibfield  {journal} {\bibinfo  {journal} {J. Stat. Mech.}\ ,\ \bibinfo
  {pages} {P12005}} (\bibinfo {year} {2012}{\natexlab{b}})}\BibitemShut
  {NoStop}%
\bibitem [{\citenamefont {Aron}\ \emph {et~al.}(2014)\citenamefont {Aron},
  \citenamefont {Barci}, \citenamefont {Cugliandolo}, \citenamefont {Arenas},\
  and\ \citenamefont {Lozano}}]{Aron2014}%
  \BibitemOpen
  \bibfield  {author} {\bibinfo {author} {\bibfnamefont {C.}~\bibnamefont
  {Aron}}, \bibinfo {author} {\bibfnamefont {D.~G.}\ \bibnamefont {Barci}},
  \bibinfo {author} {\bibfnamefont {L.~F.}\ \bibnamefont {Cugliandolo}},
  \bibinfo {author} {\bibfnamefont {Z.~G.}\ \bibnamefont {Arenas}}, \ and\
  \bibinfo {author} {\bibfnamefont {G.~S.}\ \bibnamefont {Lozano}},\ }\href
  {http://stacks.iop.org/1742-5468/2014/i=9/a=P09008} {\bibfield  {journal}
  {\bibinfo  {journal} {J. Stat. Mech.}\ ,\ \bibinfo {pages} {P09008}}
  (\bibinfo {year} {2014})}\BibitemShut {NoStop}%
\bibitem [{\citenamefont {H{\"a}nggi}(1978)}]{Hanggi1978}%
  \BibitemOpen
  \bibfield  {author} {\bibinfo {author} {\bibfnamefont {P.}~\bibnamefont
  {H{\"a}nggi}},\ }\href {\doibase 10.5169/seals-114941} {\bibfield  {journal}
  {\bibinfo  {journal} {Helv. Phys. Acta}\ }\textbf {\bibinfo {volume} {51}},\
  \bibinfo {pages} {183} (\bibinfo {year} {1978})}\BibitemShut {NoStop}%
\bibitem [{\citenamefont {H{\"a}nggi}(1980)}]{Hanggi1980}%
  \BibitemOpen
  \bibfield  {author} {\bibinfo {author} {\bibfnamefont {P.}~\bibnamefont
  {H{\"a}nggi}},\ }\href {\doibase 10.5169/seals-115133} {\bibfield  {journal}
  {\bibinfo  {journal} {Helv. Phys. Acta}\ }\textbf {\bibinfo {volume} {53}},\
  \bibinfo {pages} {491} (\bibinfo {year} {1980})}\BibitemShut {NoStop}%
\bibitem [{\citenamefont {H{\"a}nggi}\ and\ \citenamefont
  {Thomas}(1982)}]{Hanggi1982}%
  \BibitemOpen
  \bibfield  {author} {\bibinfo {author} {\bibfnamefont {P.}~\bibnamefont
  {H{\"a}nggi}}\ and\ \bibinfo {author} {\bibfnamefont {H.}~\bibnamefont
  {Thomas}},\ }\href@noop {} {\bibfield  {journal} {\bibinfo  {journal} {Phys.
  Rep.}\ }\textbf {\bibinfo {volume} {88}},\ \bibinfo {pages} {207} (\bibinfo
  {year} {1982})}\BibitemShut {NoStop}%
\bibitem [{\citenamefont {Klimontovich}(1994)}]{Klimontovich}%
  \BibitemOpen
  \bibfield  {author} {\bibinfo {author} {\bibfnamefont {Y.~L.}\ \bibnamefont
  {Klimontovich}},\ }\href {\doibase 10.1070/PU1994v037n08ABEH000038}
  {\bibfield  {journal} {\bibinfo  {journal} {Physics-Uspekhi}\ }\textbf
  {\bibinfo {volume} {37}},\ \bibinfo {pages} {737} (\bibinfo {year}
  {1994})}\BibitemShut {NoStop}%
\bibitem [{\citenamefont {Haussmann}\ and\ \citenamefont
  {Pardoux}(1986)}]{Haussmann1986}%
  \BibitemOpen
  \bibfield  {author} {\bibinfo {author} {\bibfnamefont {U.~G.}\ \bibnamefont
  {Haussmann}}\ and\ \bibinfo {author} {\bibfnamefont {E.}~\bibnamefont
  {Pardoux}},\ }\href
  {http://www.jstor.org/discover/10.2307/2243859?uid=2&uid=4&sid=21105181922903}
  {\bibfield  {journal} {\bibinfo  {journal} {Ann. Probab.}\ }\textbf {\bibinfo
  {volume} {14}},\ \bibinfo {pages} {1188} (\bibinfo {year}
  {1986})}\BibitemShut {NoStop}%
\bibitem [{\citenamefont {Millet}\ \emph {et~al.}(1989)\citenamefont {Millet},
  \citenamefont {Nualart},\ and\ \citenamefont {Sanz}}]{Millet1989}%
  \BibitemOpen
  \bibfield  {author} {\bibinfo {author} {\bibfnamefont {A.}~\bibnamefont
  {Millet}}, \bibinfo {author} {\bibfnamefont {D.}~\bibnamefont {Nualart}}, \
  and\ \bibinfo {author} {\bibfnamefont {M.}~\bibnamefont {Sanz}},\ }\href
  {http://www.jstor.org/discover/10.2307/2244207?uid=2&uid=4&sid=21105181922903}
  {\bibfield  {journal} {\bibinfo  {journal} {Ann. Probab.}\ }\textbf {\bibinfo
  {volume} {17}},\ \bibinfo {pages} {208} (\bibinfo {year} {1989})}\BibitemShut
  {NoStop}%
\bibitem [{\citenamefont {Lau}\ and\ \citenamefont
  {Lubensky}(2007)}]{Lubensky2007}%
  \BibitemOpen
  \bibfield  {author} {\bibinfo {author} {\bibfnamefont {A.~W.~C.}\
  \bibnamefont {Lau}}\ and\ \bibinfo {author} {\bibfnamefont {T.~C.}\
  \bibnamefont {Lubensky}},\ }\href {\doibase 10.1103/PhysRevE.76.011123}
  {\bibfield  {journal} {\bibinfo  {journal} {Phys. Rev. E}\ }\textbf {\bibinfo
  {volume} {76}},\ \bibinfo {pages} {011123} (\bibinfo {year}
  {2007})}\BibitemShut {NoStop}%
\bibitem [{\citenamefont {Zinn-Justin}(2002)}]{Zinn-Justin}%
  \BibitemOpen
  \bibfield  {author} {\bibinfo {author} {\bibfnamefont {J.}~\bibnamefont
  {Zinn-Justin}},\ }\href@noop {} {\emph {\bibinfo {title} {Quantum field
  theory and critical phenomena}}}\ (\bibinfo  {publisher} {Oxford University
  Press},\ \bibinfo {address} {USA},\ \bibinfo {year} {2002})\BibitemShut
  {NoStop}%
\bibitem [{\citenamefont {Volpe}\ \emph {et~al.}(2010)\citenamefont {Volpe},
  \citenamefont {Helden}, \citenamefont {Brettschneider}, \citenamefont
  {Wehr},\ and\ \citenamefont {Bechinger}}]{Volpe2010}%
  \BibitemOpen
  \bibfield  {author} {\bibinfo {author} {\bibfnamefont {G.}~\bibnamefont
  {Volpe}}, \bibinfo {author} {\bibfnamefont {L.}~\bibnamefont {Helden}},
  \bibinfo {author} {\bibfnamefont {T.}~\bibnamefont {Brettschneider}},
  \bibinfo {author} {\bibfnamefont {J.}~\bibnamefont {Wehr}}, \ and\ \bibinfo
  {author} {\bibfnamefont {C.}~\bibnamefont {Bechinger}},\ }\href {\doibase
  10.1103/PhysRevLett.104.170602} {\bibfield  {journal} {\bibinfo  {journal}
  {Phys. Rev. Lett.}\ }\textbf {\bibinfo {volume} {104}},\ \bibinfo {pages}
  {170602} (\bibinfo {year} {2010})}\BibitemShut {NoStop}%
\bibitem [{\citenamefont {Brenner}(1961)}]{Brenner1961}%
  \BibitemOpen
  \bibfield  {author} {\bibinfo {author} {\bibfnamefont {H.}~\bibnamefont
  {Brenner}},\ }\href@noop {} {\bibfield  {journal} {\bibinfo  {journal} {Chem.
  Eng. Sci.}\ }\textbf {\bibinfo {volume} {16}},\ \bibinfo {pages} {242}
  (\bibinfo {year} {1961})}\BibitemShut {NoStop}%
\bibitem [{\citenamefont {van Kampen}(1981)}]{vanKampenIto}%
  \BibitemOpen
  \bibfield  {author} {\bibinfo {author} {\bibfnamefont {N.}~\bibnamefont {van
  Kampen}},\ }\href {\doibase 10.1007/BF01007642} {\bibfield  {journal}
  {\bibinfo  {journal} {Journal of Statistical Physics}\ }\textbf {\bibinfo
  {volume} {24}},\ \bibinfo {pages} {175} (\bibinfo {year} {1981})}\BibitemShut
  {NoStop}%
\bibitem [{\citenamefont {Fox}(1978)}]{Fox1978}%
  \BibitemOpen
  \bibfield  {author} {\bibinfo {author} {\bibfnamefont {R.~F.}\ \bibnamefont
  {Fox}},\ }\href {\doibase http://dx.doi.org/10.1016/0370-1573(78)90145-X}
  {\bibfield  {journal} {\bibinfo  {journal} {Physics Reports}\ }\textbf
  {\bibinfo {volume} {48}},\ \bibinfo {pages} {179 } (\bibinfo {year}
  {1978})}\BibitemShut {NoStop}%
\bibitem [{\citenamefont {Brown}(1963)}]{Brown1963}%
  \BibitemOpen
  \bibfield  {author} {\bibinfo {author} {\bibfnamefont {W.~F.}\ \bibnamefont
  {Brown}},\ }\href@noop {} {\bibfield  {journal} {\bibinfo  {journal} {Phys.
  Rev.}\ }\textbf {\bibinfo {volume} {130}},\ \bibinfo {pages} {1677} (\bibinfo
  {year} {1963})}\BibitemShut {NoStop}%
\bibitem [{\citenamefont {Landau}\ and\ \citenamefont
  {Lifshitz}(1935)}]{Landau1935}%
  \BibitemOpen
  \bibfield  {author} {\bibinfo {author} {\bibfnamefont {L.~D.}\ \bibnamefont
  {Landau}}\ and\ \bibinfo {author} {\bibfnamefont {E.~M.}\ \bibnamefont
  {Lifshitz}},\ }\href@noop {} {\bibfield  {journal} {\bibinfo  {journal}
  {Phys. Z. Sowjetunion}\ }\textbf {\bibinfo {volume} {8}},\ \bibinfo {pages}
  {153} (\bibinfo {year} {1935})}\BibitemShut {NoStop}%
\bibitem [{\citenamefont {Gilbert}(1955)}]{Gilbert1955}%
  \BibitemOpen
  \bibfield  {author} {\bibinfo {author} {\bibfnamefont {T.~L.}\ \bibnamefont
  {Gilbert}},\ }\href@noop {} {\bibfield  {journal} {\bibinfo  {journal} {Phys.
  Rev.}\ }\textbf {\bibinfo {volume} {100}},\ \bibinfo {pages} {1243} (\bibinfo
  {year} {1955})}\BibitemShut {NoStop}%
\bibitem [{\citenamefont {Gilbert}(2004)}]{Gilbert2004}%
  \BibitemOpen
  \bibfield  {author} {\bibinfo {author} {\bibfnamefont {T.~L.}\ \bibnamefont
  {Gilbert}},\ }\href@noop {} {\bibfield  {journal} {\bibinfo  {journal} {IEEE
  Trans. Magn.}\ }\textbf {\bibinfo {volume} {40}},\ \bibinfo {pages} {3443}
  (\bibinfo {year} {2004})}\BibitemShut {NoStop}%
\bibitem [{\citenamefont {Rom\'a}\ \emph {et~al.}(2014)\citenamefont {Rom\'a},
  \citenamefont {Cugliandolo},\ and\ \citenamefont {Lozano}}]{Roma2014}%
  \BibitemOpen
  \bibfield  {author} {\bibinfo {author} {\bibfnamefont {F.}~\bibnamefont
  {Rom\'a}}, \bibinfo {author} {\bibfnamefont {L.~F.}\ \bibnamefont
  {Cugliandolo}}, \ and\ \bibinfo {author} {\bibfnamefont {G.~S.}\ \bibnamefont
  {Lozano}},\ }\href {\doibase 10.1103/PhysRevE.90.023203} {\bibfield
  {journal} {\bibinfo  {journal} {Phys. Rev. E}\ }\textbf {\bibinfo {volume}
  {90}},\ \bibinfo {pages} {023203} (\bibinfo {year} {2014})}\BibitemShut
  {NoStop}%
\end{thebibliography}
%

\end{document}